\documentclass[showpacs,twocolumn,floatfix,amsmath,amssymb,superscriptaddress,prb,nopacs]{revtex4-1}

\pdfoutput=1

\usepackage{dcolumn}
\usepackage[dvipdfmx]{graphicx}
\usepackage{mathrsfs}
\usepackage{accents}
\usepackage{bm}
\usepackage{dsfont}
\usepackage{color}
\usepackage{here}
\usepackage{longtable}

\makeatletter
\def\btt#1{\texttt{\@backslashchar#1}}
\DeclareRobustCommand\bblash{\btt{\@backslashchar}} \makeatother

\newcommand{\bea}{\begin{eqnarray}}
\newcommand{\eea}{\end{eqnarray}}
\newcommand{\bt}{\textbf}

\newcommand{\phd}{\phantom{\dag}}
\newcommand{\ph}{\phantom{.}}

\newcommand{\noi}{\noindent}
\newcommand{\no}{\nonumber}

\begin{document}

\title{Nodal Andreev Spectra in Multi-Majorana Three-Terminal Josephson Junctions}

\author{Keimei Sakurai} 
\affiliation{Department of Applied Physics, Hokkaido University, Sapporo 060-8628, Japan}

\author{Maria Teresa Mercaldo} 
\affiliation{Dipartimento di Fisica ``E. R. Caianiello", Universit\`a di Salerno, IT-84084 Fisciano (SA), Italy}

\author{Shingo Kobayashi}
\affiliation{Department of Applied Physics, Nagoya University, Nagoya, 464-8603, Japan}
\affiliation{Institute for Advanced Research, Nagoya University, Nagoya 464-8601, Japan} 
 
\author{Ai Yamakage}
\affiliation{Department of Physics, Nagoya University, Nagoya 464-8602, Japan}
 
\author{Satoshi Ikegaya} 
\affiliation{Max-Planck fur Festkorperforschung, Heisenbergstrasse 1, D-70569 Stuttgart, Germany}

\author{Tetsuro Habe} 
\affiliation{Department of Applied Physics, Hokkaido University, Sapporo 060-8628, Japan}

\author{Panagiotis Kotetes} 
\affiliation{CAS Key Laboratory of Theoretical Physics, Institute of Theoretical Physics, Chinese Academy of Sciences, Beijing 100190, China} 

\author{Mario Cuoco} 
\affiliation{CNR-SPIN, I-84084 Fisciano (Salerno), Italy}
\affiliation{Dipartimento di Fisica ``E. R. Caianiello", Universit\`a di Salerno, IT-84084 Fisciano (SA), Italy}

\author{Yasuhiro Asano}
\affiliation{Department of Applied Physics, Hokkaido University, Sapporo 060-8628, Japan}

\begin{abstract}
We investigate the Andreev-bound-state (ABS) spectra of three-terminal Josephson junctions which consist of 1D topological superconductors (TSCs) harboring multiple zero-energy edge Majorana bound states (MBSs) protected by chiral symmetry. Our theoretical analysis relies on the exact numerical diagonalization of the Bogoliubov-de Gennes (BdG) Hamiltonian descri\-bing the three inter\-fa\-ced TSCs, complemented by an effective low-energy description solely based on the coupling of the interfacial MBSs arising before the leads get contacted. Considering the 2D synthetic space spanned by the two independent superconducting phase dif\-fe\-ren\-ces, we demonstrate that the ABS spectra may contain either point or line nodes, and identify $\mathbb{Z}_2$ topological invariants to classify them. We show that the resulting type of nodes depends on the number of preexisting interfacial MBSs, with nodal lines necessarily appearing when two TSCs harbor an unequal number of MBSs. Specifically, the precise number of interfacial MBSs determines the periodicity of the spectrum under $2\pi$-slidings of the phase dif\-fe\-ren\-ces and, as a result, also controls the shape of the nodal lines in synthetic space. When chiral symmetry is preserved, the lines are open and coincide with high-symmetry lines of synthetic space, while when it is violated the lines can also transform into loops and chains. The nodal spectra are robust by virtue of the inherent particle-hole symmetry of the BdG Hamiltonian, and give rise to distinctive experimental signatures that we identify. 
\end{abstract}
 
\maketitle

\section{Introduction}

Spin-triplet superconductors (SCs) are marked by Cooper pairs possessing an odd-parity orbital configuration and an exchange-symmetric spin-$1$ angular momentum~\cite{Sigrist,TanakaPWave,ReadGreen,Ivanov,Kitaev01,VolovikBook,Maeno2012,SatoAndo}. The former aspect is the source of anomalous proximity effects \cite{Tanaka04,Asano06}, while the latter is particularly attractive for implementing cutting-edge functionalities, including the coherent control and manipulation of the Cooper pairs. Due to the intrinsic coupling between spin and orbital degrees of freedom, spin-triplet SCs exhibit nonstandard response to Zeeman/ferromagnetic fields~\cite{Murakami,Dumi1,Dumi2,Wright,Mercaldo2016,Mercaldo2017,Mercaldo2018}, spin-sensitive Josephson transport~\cite{Sengupta,Yakovenko,Yakovenko1,Cuoco1,Cuoco2,BrydonJunction,Gentile,Brydon1,Brydon2}, and pave the way to energy efficient superconducting spintronics~\cite{Linder2015}. Even more, p-wave SCs constitute the prototypical topological superconductors (TSCs) harboring protected zero-energy modes, the so-called Majorana bound states (MBSs), whose manipulation open perspectives for topological quantum computing~\cite{Ivanov,Kitaev03,Nayak,Alicea}.

Although spin-triplet pairing is less common than the conventional spin-singlet one, experimental observations in the last decades have gathered many evidences for spin-triplet superconductivity in a large variety of materials, such as heavy-fermion compounds~\cite{Stewart,Saxena,Kyogaku}, 
non-centrosymmetric materials~\cite{Bauer,Nishiyama}, organic conductors~\cite{Lebed,Lee,Shinagawa}, layered oxides~\cite{Ishida}, doped topological insulators~\cite{Matano}, and more recently in the Cr-based pnictide K$_2$Cr$_3$As$_3$~\cite{Bao15,Cuono19}. In addition to intrinsic materials, artificial spin-triplet SCs can be engineered in a large variety of quantum-material and -device platforms based on conventional spin-singlet pairing. For instance, an established route for achie\-ving spin singlet-triplet pair conversion is found in he\-te\-ro\-struc\-tu\-res consisting of spin-singlet SCs interfaced with magnetically-active materials~\cite{Buzdin,Bergeret,Eschrig,Linder2015}. Furthermore, p-wave spin-triplet pairing emerges effectively in to\-po\-lo\-gi\-cal insulators proximity-coupled to conventional SCs~\cite{FuKane,KlinovajaKramersPairs}, semiconductor-superconductor hybrids~\cite{SauPRL,AliceaPRB,LutchynPRL,OregPRL,Jay,Jens,Haim,KlinovajaTRIparafermions,KotetesNoZeeman,Hell,PientkaPlanar,HaimReview,KotetesPRL2019}, and magnetic atomic chains on SCs~\cite{Choy,NadgPerge,Nakosai,Braunecker,Klinovaja,Vazifeh,Pientka,Ojanen1,Heimes,Brydon,Li,Heimes2,JinAn,Silas,Andolina}. 

In this framework, Josephson junctions are central for the engineering of Andreev bound states (ABSs) and the design of topological quantum-computing architectures based on MBSs~\cite{Nayak,Ivanov,Alicea,Sau111,Heck}. For instance, in junctions con\-si\-sting of two TSCs with a phase difference $\Delta \phi$, coupling the interface MBSs can yield 4$\pi$-periodic ABS ener\-gy dispersions with a crossing at $\Delta\phi=\pi$ which is protected by the conservation of fermion parity (FP)~\cite{FuKaneJ,CarloJ}. Such a behavior becomes significantly modified when con\-si\-de\-ring a topological junction with multiple MBSs at each end protected by a chiral symmetry~\cite{Altland,Schnyder,KitaevClassi,Ryu,TeoKane,KotetesClassi,Shiozaki,Chiu}. There, one can achieve a nonstandard control for the ABS spectra~\cite{Mercaldo2019}, e.g. through electrical gating, and obtain effective electronic structures with nodes or Weyl points in synthetic space that allow to realize fermion-parity pumping in a suitably-designed cycle~\cite{KotetesPRL2019}. 

Alternatively, engineering topologically-nontrivial ABS dispersions with nodal and Weyl points is also possible in three and four-terminal Josephson junctions consisting of conventional s-wave SCs connected through a normal-metal region~\cite{vanHeck14,Yokoyama15,Riwar16}. In this case, the presence of the Weyl points and their associated topological monopole charge leads to a quantized transconductance~\cite{Riwar16,Eriksson17}, a remarkable finding that persists even in the presence of spin-orbit coupling~\cite{Yokoyama15} or an external magnetic field~\cite{Meyer17,Xie17}. Along these lines, the substitution of the conventional SC leads by topological ones provides a new twist in the ABS engi\-nee\-ring. This was recently proposed in Ref.~\onlinecite{Houzet18} for four-terminal Josephson junctions consisting of 1D TSCs with a single MBS per edge. Such junctions have been shown to exhibit finite-energy Weyl crossings between the two lowest ABSs. One may thus ask whether the employment of TSCs can lead to novel effects compared to the case of junctions building upon s-wave SCs, especially because the conventional classification of topological semimetallic phases does not directly apply to multi-terminal supercon\-ducting setups. Indeed, while in a crystal the existence of Weyl nodes requires the breaking of time or/and inversion symmetry, in four-terminal junctions Weyl points of ABSs can also occur in a time-reversal symmetric scenario.

In this work, we investigate the emergence and design of topologically-nontrivial ABS bands in multi-terminal Josephson contacts consisting of 1D TSCs which accommodate multiple MBSs per edge. Such multi-MBS topological scenarios become accessible in both intrinsic~\cite{Dumi1,Dumi2,Mercaldo2016,Mercaldo2017,Mercaldo2018,Mercaldo2019} or effective~\cite{Jay,Jens,Haim,KlinovajaTRIparafermions,KotetesNoZeeman,Hell,PientkaPlanar,HaimReview,KotetesPRL2019,Heimes2,JinAn,Silas,Andolina} p-wave spin-triplet SCs, with the chiral symmetry being associated with a Kramers degeneracy or a sublattice symmetry. Consi\-de\-ring the 2D synthetic space of the two independent phase differences appearing between the supercon\-ducting leads of a three-terminal junction, we demonstrate that the ABS dispersions feature point nodes or nodal lines which depend on the number of MBSs occurring at the junction's interface. This nodal structure is a distinct fingerprint of the topological character of the ABSs, when these are constructed by means of coupled edge MBSs. To transparently expose the properties of the emergent ABS spectra, we employ a real-space nu\-me\-ri\-cal analysis relying on the evaluation of the Bogoliubov - de Gennes (BdG) energy spectra of the spin-triplet SCs, combined with an effective low-energy description that relies on the hybridization of the MBSs localized at the intersection of the three-terminal junction. 

We find that the topological character of the ABS spectra depends on the MBS configuration. Specifically, a nodal line spectrum is always obtained in cases where two TSCs possess an unequal number of MBSs. We further show that the geometrical structure of the nodal lines in synthetic space depends on the invariance of the BdG Hamiltonian under $2\pi$-sli\-dings of the phase differences between the superconducting leads. This becomes particularly transparent in the low-energy MBSs description that we employ. The MBS Hamiltonian obtained in this low-energy description is bound to specific constraints stemming from the $2\pi$-sliding symmetries. When chiral symmetry is present, these constraints allow us to predict the location of the nodal points and lines in synthetic space, as well as to define related $\mathbb{Z}_2$ indices which reflect their topological robustness. Specifically, chiral symmetry introduces high-symmetry points and lines in synthetic space, where the nodal points and open lines are set to appear as a result of the sliding symmetries. Even more, the conservation of FP and the inherent particle-hole symmetry (PHS) of the MBS Hamiltonian, gua\-ran\-tee that the nodal lines are robust even when chiral symmetry becomes violated. The nodal lines can be only removed either when a pair of identical open lines meet and annihilate, or when a line-topology conversion takes place, upon which, the open lines evolve into loops or chains which can be continuously deformed into points and subsequently disappear. 

It is quite remarkable that a number of topologically-nontrivial band structures which were only recently discovered in real materials, e.g. nodal loops and nodal lines semimetals~\cite{Liu,Liu1,Neupane,Bian,Wu,Wang,Yan}, can be engineered and studied in a synthetic space using the multi-terminal Josephson junctions of TSCs. Even more importantly, the nodal spectra tailored in such TSC devices are protected against weak disorder by means of FP conservation and PHS. This additionally implies that by adiabatically and selectively sweeping suitable paths in synthetic space, cf. Ref.~\onlinecite{KotetesPRL2019}, a number of disorder-robust Josephson transport phenomena can be experimentally accessed, which reflect the topological character of the ABS spectra and the underlying presence of MBSs.

The paper is organized as follows. In Sec.~II we describe the here-considered three-terminal junction consisting of p-wave topological SC chains. In Sec.~III we discuss the chiral symmetry of the Hamiltonian for each superconducting lead, and derive an effective low-energy description based on the interface MBSs. In Sec.~IV, we demonstrate the main features of the ABS spectra with respect to the number of MBSs at the SCs' intersection, in terms of the arising differences between the phases of the SCs. In Sec.~V we present the symmetry aspects of the low-energy model and unveil the to\-po\-lo\-gi\-cal character of the nodal points and lines in the ABSs spectra. Section~VI focuses on the hybridization of MBSs on a given TSC, the concomitant robustness of the point and line nodes, as well as the emergence of nodal loops and chains in synthetic space. Further, we discuss two possible physical situations leading to this hybridization, i.e. chiral-symmetry breaking and the presence of trivial ABSs, and propose experimental stategies to disentangle them. In Sec.~VII we discuss a number of experimental Josepshon transport properties that allow identifying the nodal ABS spectra. Our conclusions are reported in Sec.~VIII.

\begin{figure*}[t!]
\begin{center}
\includegraphics[width=0.85\textwidth]{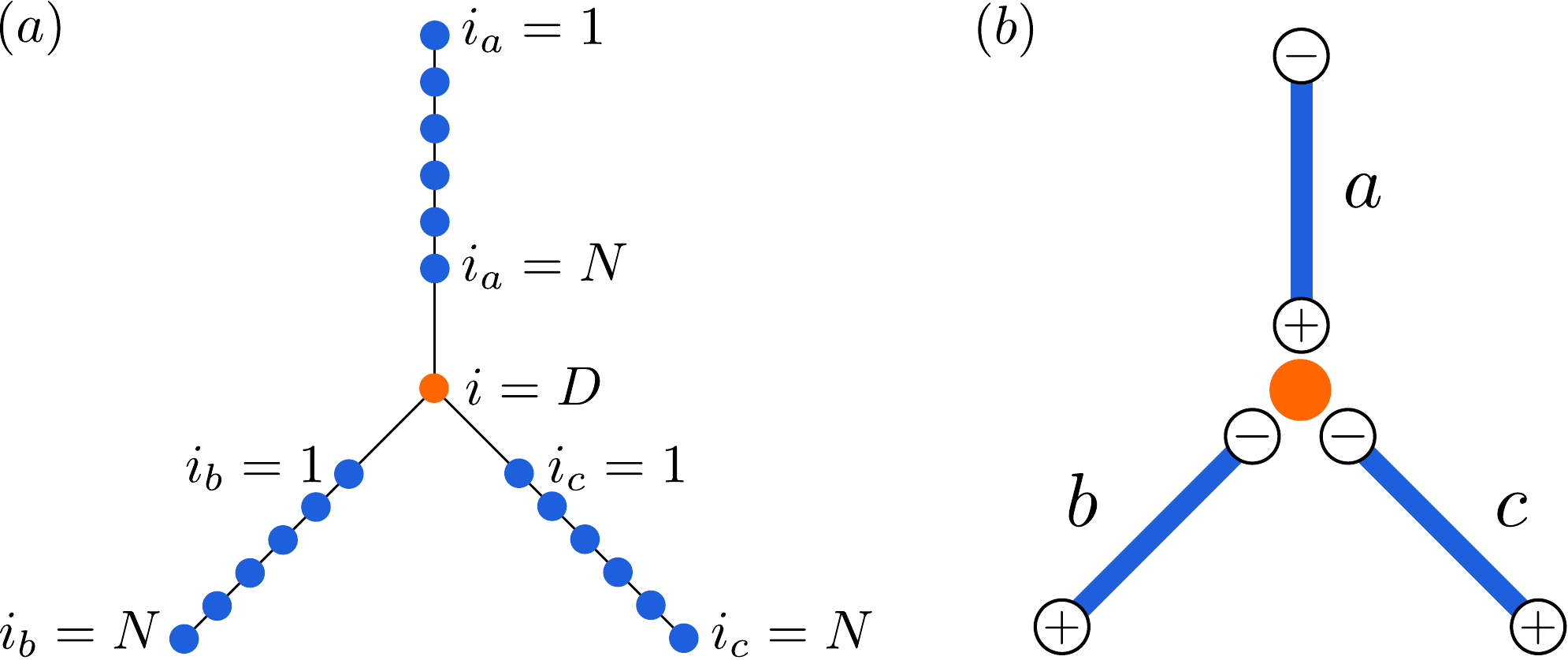}
\caption{(a) Schematic picture of the three-terminal junction consisting of three 1D topological superconductors (TSCs), each one of which is described by a single finite-sized chain (blue circles). The superconducting leads are connected in an indirect manner, through a single site (labelled as $i=D$). (b) depicts the configuration of chiralities dictating the eigenstates of the zero-energy Majorana bound states (MBSs) appearing at the edges of the three TSCs for $\phi_a=\phi_b=\phi_c=0$. All the MBSs constitute eigenstates of the chiral-symmetry operator $\hat{\Gamma}(0)$ defined for zero phase differences between the superconducting leads. Here, $\pm$ indicates the chirality of the edge MBSs. We remark that the specific choice made for the chiralities of the MBSs does not affect our results, but only sets the location of the high-symmetry points and lines in synthetic space.}
\label{fig:model}
\end{center}
\end{figure*}

\section{Model and methodology}

In this section, we introduce the setup and model for the here-considered three-terminal junction con\-si\-sting of p-wave topological SC chains, and briefly describe the numerical approach to be followed. At this point, we remind the reader that, while in the following we consider a concrete model, the qualitative picture drawn from our results has a general character. In this sense, our work is not only applicable to the intrinsic p-wave SCs discussed here, but it also extends to engineered TSCs based on superconductor-semiconductor hybrids~\cite{Jay,Jens,Haim,KlinovajaTRIparafermions,KotetesNoZeeman,Hell,PientkaPlanar,HaimReview,KotetesPRL2019,Heimes2,JinAn,Silas,Andolina} and topological magnetic chains~\cite{Heimes2,JinAn,Silas,Andolina}, where the multiple MBSs at each TSC edge arise from the presence of either a Kramers degeneracy or a sublattice symmetry.

\subsection{Model}

We consider a three-terminal (Y-shaped) junction con\-si\-sting of three 1D TSCs labelled by $\alpha=(a,b,c)$, as shown in Fig.~\ref{fig:model}. Each TSC has a length of $N$ sites. The three TSCs are electronically connected through their individual coupling to a normal conducting region that consists of a noninteracting quantum dot with a single spin-degenerate level with a chemical potential $\mu_D$. The quantum dot is considered to be tunnel-coupled to the superconducting leads with a charge-transfer strength $t_D$. Each one of the TSCs hosts an even integer number of MBSs, where half of them are localized at the edge near the junction's interface and the remaining appear on the remaining edge, which is far away from the intersection of the superconducting leads. We assume that the length of each SC is sufficiently large, so that the MBSs away from the junction's interface become irrelevant and thus can be neglected.

Each 1D TSC lead is described by a standard spinful p-wave BCS-type mean-field Hamiltonian $\hat{H}_\alpha$ in the additional presence of an external Zeeman field, cf  Ref.~\onlinecite{Mercaldo2019}:
\bea
\hat{H}_{\alpha}&=&-t\sum_{i_\alpha=1}^{N-1}\sum_{s=\{\uparrow,\downarrow\}}\big(c_{i_\alpha+1,s}^\dag c_{i_\alpha,s}+c_{i_\alpha,s}^\dag c_{i_\alpha+1,s}\big)\no\\
&+&\sum_{i_\alpha=1}^{N} \sum_{s,s'=\{\uparrow,\downarrow\}}
c_{i_\alpha, s}^\dag\big(-\mu_\alpha\delta_{s,s'}-\bm{h}_\alpha\cdot\hat{\bm{\sigma}}_{s,s'}\big)c_{i_\alpha,s'}\no\\
&+&\sum_{i_\alpha=1}^{N-1}\sum_{s=\{\uparrow,\downarrow\}}\Big[\Delta_s(i_\alpha)e^{i\phi_\alpha}c_{i_\alpha+1,s}^\dag c_{i_\alpha,s}^\dag+{\text{h.c.}}\Big]\,.
\label{pwaveHamiltonian}
\eea

\noi For a schematic representation of the Hamiltonian see Fig.~\ref{fig:model}(a). In the above, $c_{i_{\alpha},s}$ is the annihilation operator of an electron at the $i$-th site of the $\alpha$-th TSC with spin $s=\{\uparrow, \downarrow\}$, and $t$ is the hopping integral between the nearest neighbor lattice sites. In the $\alpha$-th TSC, $\mu_\alpha$ is the chemical potential, $\bm{h}_\alpha=h_\alpha(\sin\theta_\alpha,0,\cos\theta_\alpha)$ represents a Zeeman field lying in $xz$ plane, and $\hat{\sigma}_{1,2,3}$ define the standard Pauli matrices in spin space. We assume that the Zeeman field is spatially uniform in each TSC. We introduce the site- and spin-dependent pair potentials $\Delta_{\uparrow,\downarrow}(i_{\alpha})$ by employing the standard decoupling of the attractive quartic interaction~\cite{Mercaldo2016}. In the case of $\Delta_\uparrow\neq\Delta_\downarrow$, the super\-con\-ducting state becomes nonunitary~\cite{Sigrist}. Throughout the present analysis we assume that the amplitude of the pair potential is either uniform or obtained by a self-consistent iterative procedure~\cite{Mercaldo2016}. Note, however, that the qualitative picture drawn from our results is independent of which choice is made.

The $\hat{H}_{a,b,c}$ are supplemented by the Hamiltonian for the quantum dot, so that the complete model reads 
\bea
\hat{H}=\sum_{\alpha=a,b,c}\hat{H}_{\alpha}+\hat{H}_{\textrm{link}}\,,\label{totalHamiltonian}
\eea

\noi where we introduced 
\bea
\hat{H}_{\textrm{link}}&=&-t_D\sum_{s=\{\uparrow,\downarrow\}}\big(c_{i_a=N,s}^\dag\,c_{D,s}+c_{i_b=1,s}^\dag\,c_{D, s}\no\\
&+&c_{i_c=1,s}^\dag\,c_{D,s}+\textrm{h.c.}\big)-\sum_{s=\{\uparrow,\downarrow\}}\mu_D\,c_{D,s}^\dag\,c_{D,s}\,.\quad
\label{h_link}
\eea

\noi In Eq.~(\ref{h_link}), $c_{D,s}$ denotes the annihilation operator of the dot electron with spin-projection $s=\uparrow,\downarrow$, while $\mu_D$ sets the energy scale for the spin-degenerate dot level. 

In the following sections, we employ the diagonalization of the BdG Hamiltonian obtained from $\hat{H}$, to investigate the number of the ingap nonzero- as well as zero-energy states (ZESs) appearing at the three-terminal junction's interface. Specifically, we are interested in the evolution of the ingap-bound-state spectra upon va\-rying the superconducting phase differences 
\bea
\phi_{\alpha\beta}=\phi_\alpha-\phi_\beta\phd{\rm with}\phd\alpha,\beta=a,b,c\label{phi_dif}
\eea

\noi defined by the phases $\phi_{a,b,c}$ characterizing each TSC. 

\subsection{Numerical methods}

For the numerical simulations we assume that all the SCs have an equal length, given by $N=200$ sites, while we set $\mu_\alpha=\mu_D=0.5t$. For the chosen windows of parameter values, small variations in the size of the superconducting leads leave our numerical results practically unaltered. For our purposes, it is sufficient to consider the amplitude of the pair potentials to be the same in each TSC. In particular, we consider $\Delta_\uparrow=\Delta_\downarrow=0.4t$ to obtain a superconducting state with two MBSs per edge, while the choice $\Delta_\uparrow=0.4t$ and $\Delta_\downarrow=0.0$ leads to a TSC with one MBS per edge. The phase with two MBSs can be also obtained by considering the nonunitary configuration~\cite{Mercaldo2016}. Moreover, in order to achieve superconducting states with one and two MBSs, we choose $h_\alpha=2.0t$ and $h_\alpha=1.0t$, respectively. The direction of the Zeeman field is fixed at $\theta_a=0.1\pi$, $\theta_b=0.2\pi$, and $\theta_c=0.3\pi$. We point out that, due to the topological character of the system, the conclusions of the analysis do not depend on the choice of the electronic parameters or the amplitude of the superconducting order parameters. It is the number of MBSs per edge in each TSC that mainly controls the structure of the observed ABS spectra. 

\section{Chiral symmetry and Majorana bound states}

Before analyzing the results of the numerical simulations for the lattice model of Eq.~\eqref{totalHamiltonian}, we discuss the chiral symmetry of the Hamiltonian describing each superconducting lead, and derive an effective low-energy Hamiltonian expressed uniquely in terms of the MBSs appearing near the interface of the three-terminal junction before the TSCs get contacted. As we show in the next section, the spectra obtained using the effective Hamiltonian reproduce satisfactorily the dependence of the low-energy ABSs spectra on the superconducting phase differences. 

\subsection{Chiral symmetry}

We start by discussing the chiral symmetry present in the Hamiltonian of a given TSC. Indeed, in the bulk case, each Hamiltonian of Eq.~\eqref{pwaveHamiltonian} can be recast in the form:
\bea
\hat{H}(k,\phi)=e^{i\phi\hat{\tau}_3/2}\hat{H}(k)e^{-i\phi\hat{\tau}_3/2}
\eea

\noi with
\begin{align}
\hat{H}(k)=\big[\epsilon(k)-h_x\hat{\sigma}_1-h_z\hat{\sigma}_3\big]\hat{\tau}_3+\big[\Delta_+(k)+\Delta_-(k)\hat{\sigma}_3\big]\hat{\tau}_1,
\label{h_op}
\end{align}

\noi where we introduced the quantities
\bea
\Delta_{\uparrow,\downarrow}(k)=\Delta_{\uparrow,\downarrow}\sin(k)\,,\quad\Delta_\pm(k)=\frac{\Delta_\uparrow(k)\pm\Delta_\downarrow(k)}{2}\,,\no\\
\epsilon(k)=-2t\cos(k)-\mu\,.\qquad\qquad\qquad
\eea

\noi In the above, $\hat{\tau}_{1,2,3}$ denote Pauli matrices defined in the particle-hole Nambu subspace.

The Hamiltonian preserves chiral symmetry, which is effected by the operator $\hat{\Gamma}(\phi)$ in the local $\phi$ frame, and is defined via the vanishing of the following anticommutation relation
\bea
\big\{\hat{H}(k,\phi),\hat{\Gamma}(\phi)\big\}=\hat{0}\,,\quad\hat{\Gamma}(\phi)\equiv e^{i\phi\hat{\tau}_3}\,\hat{\tau}_2\,.
\eea

\noi Since $\hat{\Gamma}^2=\hat{1}$, its eigenvalues $\Gamma$, take the va\-lues $\Gamma=\pm 1$. Due to chiral symmetry, the supercon\-duc\-ting state can be topologically characterized by intro\-ducing the following winding number
\bea
w=\frac{i}{4\pi}\int_{-\pi}^{\pi}dk\ph\textrm{Tr}\Big[\hat{\Gamma}(0)\,\hat{H}^{-1}(k)\,\partial_k\,\hat{H}(k)\Big]\in\mathbb{Z}\,.
\eea

\noi According to the bulk-boundary correspondence principle~\cite{Altland,Schnyder,KitaevClassi,Ryu}, the above implies that a number of $|w|$ MBSs will appear on each edge of the TSC. 

On general grounds, the eigenstates of any given Hamiltonian possessing chiral symmetry exhibit the following characteristic features (cf.~Refs.~\onlinecite{sato11,ikegaya15,ikegaya16}): (i) the ZESs of the BdG Hamiltonian $\hat{H}(k,\phi)$ are simultaneously eigenstates of the chiral-symmetry operator $\hat{\Gamma}$ and (ii) the nonzero energy states of $\hat{H}(k,\phi)$ are described by li\-near combinations of two states $\chi_{\pm}$ asso\-cia\-ted with the chirality eigenvalues $\Gamma=\pm1$, respectively. Namely, any nonzero energy state can be expressed as $\varphi_{E\neq0}=c_+\chi_++c_-\chi_-$, where the relation $|c_+|=|c_-|$ always holds. The way in which the MBSs become hybridized at the intersecting point of the junction can be discussed and addressed by means of the above two ge\-ne\-ral properties. 

At $\phi=0$, the two vectors 
\bea
f_{+,\uparrow}=\frac{1}{\sqrt{2}}(1,0,i,0)^{\intercal},\quad f_{+,\downarrow}=\frac{1}{\sqrt{2}}(0,1,0,i)^{\intercal},\quad\label{f_plus}
\eea

\noi span the positive-eigenvalue subspace of $\hat{\Gamma}(0)$, whereas
\bea
f_{-,\uparrow}=\frac{1}{\sqrt{2}}(i,0,1,0)^{\intercal},\quad f_{-,\downarrow}=\frac{1}{\sqrt{2}}(0,i,0,1)^{\intercal},\quad\label{f_minus}
\eea

\noi span the negative-eigenvalue sector of $\hat{\Gamma}(0)$. Here, $^{\intercal}$ effects matrix transposition. Since $\big[\hat{\Gamma}(\phi),\hat{\sigma}_3\big]=\hat{0}$, the ZESs are classified into two kinds, depending on their eigenvalues under the action of $\hat{\sigma}_3$. In particular, the vectors $f_{\pm,\uparrow}$ and $f_{\pm,\downarrow}$ belong to the positive and negative eigenvalue sectors of $\hat{\sigma}_3$, respectively. 

Besides the spin structure, it is worth commenting on the behavior of the MBS eigenstates under shifts of the superconducting phase. Specifically, a $\pi$-shift in the superconducting phase of a given TSC inverts the chirality configuration on the two edges, since $\hat{\Gamma}(\phi+\pi)=-\hat{\Gamma}(\phi)$. For instance, one has: 
\bea
e^{i\pi\hat{\tau}_3/2}f_{+,s}=f_{-,s}\phd{\rm with}\phd s=\{\uparrow,\downarrow\}\,.\label{pi_shift}
\eea

\begin{figure*}[t!]
\begin{center}
\includegraphics[width=17.0cm]{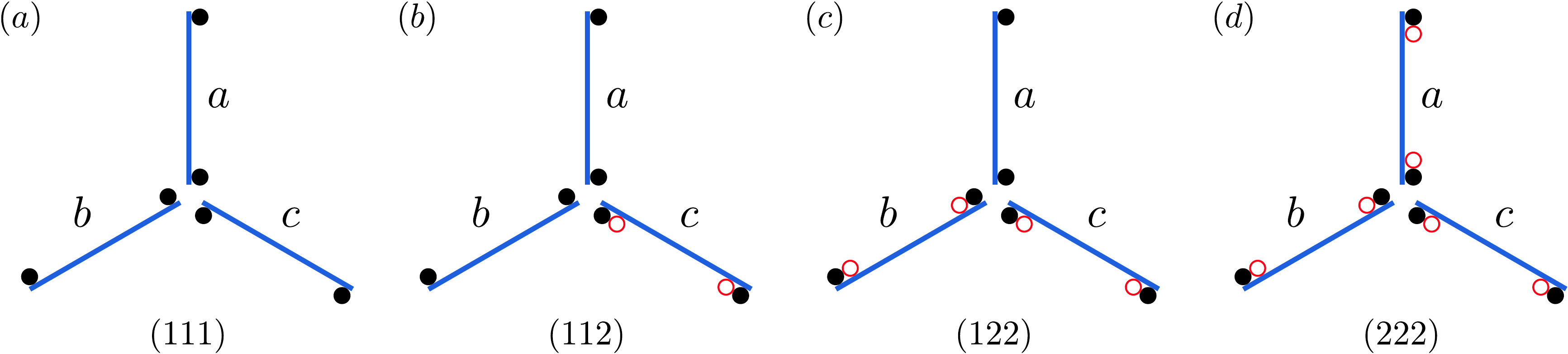}
\caption{Schematic configuration of Majorana bound states in the case of vanishing coupling between the superconducting leads and the dot, i.e. $t_D=0$. The notation $(w_a w_b w_c)$-junction indicates a three-terminal junction of the corresponding $a$-, $b$- and $c$- topological superconductor.}
\label{fig:config}
\end{center}
\end{figure*}

\noi On the other hand, a $2\pi$-shift in the superconducting phase does not modify the chirality of the eigenstates. This is in fact expected, since all physical observables should be invariant under such a shift, as a consequence of the $2\pi$-periodicity of the BdG Hamiltonian with respect to the superconducting phase. However, the MBS eigenvectors do not need to be invariant. In fact, a $2\pi$-shift introduces an overall factor of $-1$ in the MBS eigenvectors, as well as in the associated MBS creation (annihilation) operators. Specifically, we have:
\bea
e^{i2\pi\hat{\tau}_3/2}f_{\pm,s}=-f_{\pm,s}\phd{\rm with}\phd s=\{\uparrow,\downarrow\}\,.\label{2pi_shift}
\eea

\noi We note that the above behavior of the MBS eigenvectors under a $2\pi$-shift symmetry transformation, imposes specific symmetry constraints on the low-energy Hamiltonian constructed from the intersectional MBSs. In Section~\ref{sec:Topology} we rely on these properties to define topological invariants for the nodal spectra.

At this stage, we proceed by obtaining the expression for the MBS state vectors in the various cases of interest. We first consider the topologically-nontrivial phase with $|w|=2$, which can be realized, for example, by choosing $\Delta_\uparrow=\Delta_\downarrow$. At $h_x=0$, the BdG Hamiltonian in Eq.~\eqref{h_op} is block-diagonalizable into two spin sectors. The win\-ding number is $|w|=1$ in each spin sector~\cite{sato11} and, thus, leads to $|w|=2$ when accounting for both. The winding-number value $|w|=2$ can be preserved even when $h_x$ couples the two spin sectors by virtue of the chiral-symmetry effected by $\hat{\Gamma}(\phi)$. At $\phi=0$ and $\Delta_\uparrow=\Delta_\downarrow$, the state vectors of the two localized MBSs on one edge read:
\bea
\psi_{+,1}&=&\cos(\theta/2)f_{+,\uparrow}+\sin(\theta/2)f_{+,\downarrow},\label{psi_p1}\\
\psi_{+,2}&=&\sin(\theta/2)f_{+,\uparrow}-\cos(\theta/2)f_{+,\downarrow},\label{psi_p2}
\eea

\noi with $\cos\theta=h_z/|\bm{h}|$ and $\sin\theta=h_x/|\bm{h}|$. Here we suppress the part encoding the spatial dependence of the MBS state vectors. For more details regarding the structure of the MBS state vectors see Appendix~\ref{app:A}. The MBSs localize spatially from the edge, within a distance given by the coherence length. Both states in Eqs.~\eqref{psi_p1} and~\eqref{psi_p2} belong to the $\Gamma=+1$ sector. The state vectors for the two localized MBSs on the remaining edge become: 
\bea
\psi_{-,1}&=&\cos(\theta/2)f_{-,\uparrow}+\sin(\theta/2)f_{-,\downarrow}\label{psi_m1},\\ 
\psi_{-,2}&=&\sin(\theta/2)f_{-,\uparrow}-\cos(\theta/2)f_{-,\downarrow},\label{psi_m2}
\eea

\noi and belong to the $\Gamma=-1$ sector. These results are a direct consequence of the property (i) mentioned earlier.
 
We now move on with identifying the MBS state vectors in the topological phase with $|w|=1$ which, as already mentioned in our introduction, can be realized in va\-rious ways. Here, we select a nonunitary pai\-ring, where the spin-$\downarrow$ normal-phase dispersion $\epsilon(k)+h_z$ lies energe\-ti\-cal\-ly much higher than the Fermi level, and only the spin $\uparrow$ band lies at the Fermi level. Hence, one can safely choose $\Delta_\downarrow=0$ accor\-ding to the self-consistent equation for the pair potentials. The MBSs at the two edges are represented by $\psi_{+,1}$ of Eq.~\eqref{psi_p1} and $\psi_{-,1}$ of Eq.~\eqref{psi_m1} at $\phi=0$. See Appendix A for further details. 

Fi\-gu\-re~\ref{fig:model}(b) illustrates the chirality configuration of the MBSs at $\phi_a=\phi_b=\phi_c=0$, where $\pm$ indicates the chirality eigenvalue. In our numerical simulations, we focus on the ABS spectra and the modification of the number of MBS at the interface of the junction as a function of the superconducting phase differences. 

\subsection{Effective low-energy Hamiltonian based on the MBSs localized near the Y-junction's interface}

Let us now consider the low-energy description of the ABSs appearing in the three-terminal device, by focusing on the hybridization of the MBSs in the vicinity of the Y-junction's interface. We employ $\gamma_{\alpha,\nu}$ to represent the operator of the $\nu$-th MBS appearing near the interface, with $\alpha=a,b,c$ and $\nu=1,2$. In the pre\-sen\-ce of a nonvanishing phase $\phi_\alpha$ for the $\alpha$-th TSC, the effective MBS coupling Hamiltonian, say between the edges of the $a$-th and $b$-th TSC, is given by 
\bea
\hat{H}_{ab}&=&\frac{1}{2}\sum_{\nu,\nu'=1,2}\gamma_{a,\nu}\psi_{+,\nu}^\dag\big(-\tilde{t}\hat{\tau}_3\big)e^{-i\phi_{ab}\hat{\tau}_3/2}\psi_{-,\nu'}\gamma_{b,\nu'}\no\\ 
&=&\frac{i}{2}\sum_{\nu,\nu'=1,2}t_{\nu\nu'}^{ab}\cos\left(\frac{\phi_{ab}}{2}\right)\,\gamma_{a,\nu}\,\gamma_{b,\nu'}\,.\label{u_ab}
\eea

\noi In the above, $t_{\nu\nu'}^{ab}$ denotes the coupling between the $\nu$-th edge MBS of the $a$-th TSC and the $\nu'$-th edge MBS of the $b$-th TSC. The vectors $\psi_{\pm,1,2}$ are given by Eqs.~\eqref{psi_p1}-\eqref{psi_m2}. The expressions for $t_{\nu\nu'}^{ab}$ are given in Appendix~\ref{app:B}. We note that the strength of the indirect tunnel coupling appearing between a pair of TSCs due to the intervention of the dot link, is here denoted $\tilde{t}$. We point out that the inter-TSC tunnel coupling preserves the chiral symmetry of each TSC, since the following holds $\big\{\tilde{t}\hat{\tau}_3,\hat{\Gamma}(\phi)\big\}=\hat{0}$.

The effective MBS coupling Hamiltonian between the edge of the $a$-th and $c$-th TSCs can be represented in a similar fashion:
\bea
\hat{H}_{ac}&=&\frac{1}{2}\sum_{\nu,\nu'=1,2}\gamma_{a,\nu}\psi_{+,\nu}^\dag\big(-\tilde{t}\hat{\tau}_3\big)e^{-i\phi_{ac}\hat{\tau}_3/2}\psi_{-,\nu'}\gamma_{c,\nu'}\no\\ 
&=&\frac{i}{2}\sum_{\nu,\nu'=1,2}t_{\nu\nu'}^{ac}\cos\left(\frac{\phi_{ac}}{2}\right)\gamma_{a,\nu}\gamma_{c,\nu'}\,.\label{u_ca}
\eea

The hopping Hamiltonian between the $b$-th and $c$-th TSC has, however, a different expression. This is because for $\phi_{b,c}=0$, the hybridization happens between two MBS possessing the same, and here negative, chiralities. This is also schema\-ti\-cal\-ly depicted in Fig.~\ref{fig:model}(b). We find:
\bea
\hat{H}_{bc}&=&\frac{1}{2}\sum_{\nu,\nu'=1,2}\gamma_{b,\nu}\psi_{-,\nu}^\dag\big(-\tilde{t}\hat{\tau}_3\big)e^{-i\phi_{bc}\hat{\tau}_3/2}\psi_{-,\nu'}\gamma_{c,\nu'}\no\\ 
&=&\frac{i}{2}\sum_{\nu,\nu'=1,2}t_{\nu\nu'}^{bc}\sin\left(\frac{\phi_{bc}}{2}\right)\gamma_{b,\nu}\gamma_{c,\nu'}\,.\label{u_bc}
\eea

\noi For $\phi_{bc}=0$, the interface MBS of the $b$-th and $c$-th TSCs belong to the same chirality sector of $\hat{\Gamma}(\phi_b)=\hat{\Gamma}(\phi_c)$ and, therefore, they do not couple to each other upon the application of chiral-symmetry preserving perturbations. Under these circumstances, all the hopping elements in $\hat{H}_{bc}$ vanish for $\phi_{bc}=0$.

In order to proceed in our discussion, it is also useful to remind the reader that the three phase differences among the superconductors satisfy the constraint $\phi_{ab}+\phi_{bc}+\phi_{ca}=0$. This holds as long as there exists no vorticity or magnetic flux trapped in the triangular area formed by the three TSCs at the intersection. Given this condition, in this work we consider $\phi_{bc}=-\left(\phi_{ab}+\phi_{ca}\right)$ with $0\leq\phi_{ab}<2\pi$ and $0\leq\phi_{ca}<2\pi$. Using the above phase choice, $\hat{H}_{bc}$ can be rewritten as
\bea
\hat{H}_{bc}=-\frac{i}{2}\sum_{\nu,\nu'=1,2}t_{\nu\nu'}^{bc}\sin\left(\frac{\phi_{ab}+\phi_{ca}}{2}\right)\gamma_{b,\nu}\gamma_{c,\nu'}\,.\quad
\eea

\noi Thus, the total low-energy effective Hamiltonian for the coupled MBSs at the interface of the Y-junction, which determines the ABS spectrum, is given by the sum $\hat{H}_{ab}+\hat{H}_{bc}+\hat{H}_{ca}$ and can be compactly expressed as:
\bea
\hat{H}_{\rm ABS}=\frac{1}{2}\sum_{\alpha,\alpha'=a,b,c}\ph\sum_{\nu,\nu'=1,2}\gamma_{\alpha,\nu}{\cal H}_{\nu,\alpha;\nu',\alpha'}\gamma_{\alpha',\nu'}\,.\quad\label{h1}
\eea

\noi The matrix coupling Hamiltonian in Eq.~(\ref{h1}) is skew symmetric, i.e. ${\cal H}_{\nu,\alpha;\nu',\alpha'}=-{\cal H}_{\nu',\alpha';\nu,\alpha}$, i.e. $\widehat{{\cal H}}^\intercal=-\widehat{{\cal H}}$. Since the latter matrix is Hermitian, we also find the relation:
\bea 
\widehat{{\cal H}}=-\widehat{{\cal H}}^*\,.\label{phs}
\eea

The above equation reflects the presence of a built-in PHS stemming from the anticommutation relations sa\-ti\-sfied by the MBS operators. As a result, the eigenvalues of $\widehat{{\cal H}}$ come in pairs $\pm E$. For $E\neq0$ ($E=0$) we find ABS (MBS) charged (self-conjugate) quasiparticle excitations. As first discussed by Kitaev in Ref.~\onlinecite{Kitaev01}, the skew-symmetric nature of $\widehat{{\cal H}}$ further allows to introduce the real skew symmetric matrix $\widehat{\cal B}=i\widehat{\cal H}$, whose Pfaffian sa\-ti\-sfies $[{\rm Pf}(\widehat{\cal B})]^2=\det(\widehat{\cal B})$. The latter relation implies that the ABS spectra satisfy $\prod_sE_s={\rm Pf}(\widehat{B})$, with $s$ an appropriate index that gua\-ran\-tees that only one of the two PHS-related eigenvalues is taken into account. Since the zeros of ${\rm Pf}(\widehat{\cal B})$ coincide with the zeros of the ABS spectra, this Pfaffian can be employed to identify the conditions under which MBSs are present.

Besides being skew-symmetric, $\widehat{{\cal H}}$ and $\widehat{\cal B}$ satisfy additional constraints stemming from the invariance of the many-body Hamiltonian operator $\hat{H}_{\rm ABS}$ under $2\pi$-shifts of the superconducting phases. In particular, Eq.~\eqref{2pi_shift} leads to the following general relation:
\bea
\hat{U}_\alpha^\dag\widehat{\cal B}(\ldots,\phi_\alpha,\ldots)\hat{U}_\alpha=\widehat{\cal B}(\ldots,\phi_\alpha+2\pi,\ldots)\,,\quad\label{eq:relUalpha}
\eea 

\noi where we considered a $2\pi$-shift of the superconducting phase $\phi_{\alpha}$, and introduced the diagonal matrix $\hat{U}_\alpha=\mathrm{diag}(\ldots,-1\ldots-1,\ldots1,\ldots)$, which introduces a factor of $-1$ in the entries related to the MBSs of the $\alpha$-th TSC. This constraint on $\widehat{\cal B}$ is crucial for the topological protection of the nodal ABS spectra. 

\section{ABS spectra for the three-terminal Y-junction}

In this section, we determine and analyze the ABS spectra as a function of the two independent phase differences between the superconducting leads, for different values of the total number of MBSs which can be accom\-modated at the interface of the three-terminal junction. We consider that the number of MBSs on a given edge of each TSC is $w_\alpha>0$ before the three SCs become tunnel-coupled via the quantum dot. For convenience, we indicate the generic configuration of the Y-junction with the string $(w_aw_bw_c)$. Since $w_{a,b,c}\in\mathbb{N}^+$, we expect that, when $w_a+w_b+w_c$ is an odd (even) number, an odd (even or zero) number of MBSs remains at the interface after introducing the coupling among the three TSCs. Specifically, we investigate the band topology of the ABS spectra and discuss the robustness of the MBSs once a va\-ria\-tion of the phases $\phi_{ab}$ and $\phi_{ca}$ is introduced for the va\-rious $(w_aw_bw_c)$ junction configurations. We specifically study all the topologically-nontrivial configurations with (111)-, (112)-, (122)-, and (222) edge MBSs as shown in Fig.~\ref{fig:config}. The different types of ABS spectra which become accessible for the three-junction system are summarized in Table~\ref{Table:ABSspectraSummary}.

We note that, given the choice made for $\phi_{bc}$, the pre\-sen\-ce of chiral symmetry imposes that all the MBS couplings vanish at the here-termed point P of the synthetic $\left(\phi_{ab},\phi_{ca}\right)$ space with coordinates $(\pi,\pi)$. In this case, P is an inversion-symmetric point, where one further finds $\phi_b=\phi_c-2\pi=\phi_a-\pi$. Therefore, a total number of $w_{a}+w_{b}+w_{c}$ MBSs should be present at the Y-junction's interface in this case, since in such a con\-fi\-gu\-ra\-tion, the MBSs at the edge of the $a$-th TSC change their chirality from $+$ to $-$ accor\-ding to Eq.~\eqref{pi_shift}. This modification implies that all the MBSs at the interface now belong to the same chirality sector, and thus persist upon the application of chiral-symmetry respecting perturbations. 

\subsection{(111)-junction}

We start by considering the Y-junction for the con\-fi\-gu\-ra\-tion $w_{a,b,c}=1$ which is depicted in Fig.~{\ref{fig:config}(a). In the absence of a coupling between the superconducting leads, three MBSs are always present at the junction. The tunnel-, and in turn the MBS-, couplings lift the degeneracy depending on the relative phase of $\phi_{ab}$ and $\phi_{ca}$. The energy eigenvalues are obtained numerically by explicitly solving the BdG equations on the real lattice, and are presented as a function of $\phi_{ab}$ and $\phi_{ca}$ in Fig.~\ref{fig:111}(a). One MBS persists independently of $\phi_{ab}$ and $\phi_{ca}$. Such a flat-band dispersion is omitted in Fig.~\ref{fig:111}(a) in order to facilitate the graphical presentation. The ABS ener\-gy spectra for the remaining states exhibit a dispersion when $\phi_{ab}$ and $\phi_{ca}$ are varied, while a single band tou\-ching appears at zero energy at the P point $(\pi,\pi)$. Fig.~\ref{fig:111}(b) shows the energy-contour plot of the results for positive energies, i.e. $E>0$. The numerically-retrieved results can be well-explained in terms of the low-energy effective Hamiltonian in Eq.~(\ref{h1}). Indeed, the Hamiltonian $\widehat{\cal H}_{111}$ possesses the following matrix form
\bea
\widehat{\cal H}_{111}&=&
\left(\begin{array}{ccc}
0&-it^{ab}_{11}C_{ab}&-it^{ca}_{11}C_{ca}\\
it^{ab}_{11}C_{ab}&0&it^{bc}_{11}S_{bc}\\
it^{ca}_{11}C_{ca}&-it^{bc}_{11}S_{bc}&0
\end{array}\right)\\\no\\
{\rm with}\phd C_{\alpha\alpha'}&=&\cos\left(\frac{\phi_{\alpha\alpha'}}{2}\right)\,\,{\rm and}\,\,\,S_{bc}=\sin\left(\frac{\phi_{ab}+\phi_{ca} }{2}\right)\,,\no
\eea

\begin{table}[t!]
\begin{center}
\begin{tabular}{|c|c|c|}
\hline
{$(w_aw_bw_c)$}&\bt{ABS Spectra}&{\bt{\# of Majorana flat bands}}\\\hline
(111)&Point nodes&$1$\\\hline
(112)&Point \& line nodes&$0$\\\hline
(122)&Point \& line nodes&$1$\\\hline
(222)&Point \& line nodes&$0$\\\hline
\end{tabular}
\caption{Summary of our results for the ABS spectra obtained for the various types of three-terminal junctions depending on the number of edge MBSs that each one of the three TSCs harbors. The multiple MBSs on a given TSC are protected by virtue of chiral symmetry.}
\label{Table:ABSspectraSummary}
\end{center}
\end{table}

\noi where we remind that the MBS couplings $t_{\nu,\nu'}^{\alpha\alpha'}$ were introduced in Eqs.~\eqref{u_ab}-\eqref{u_bc} and their detailed expressions are found in Appendix~\ref{app:B}. We note that these depend also on the relative angle of the Zeeman fields as discussed in Appendix~\ref{app:B}, because the orientation of the Zeeman field controls the spin configuration of the MBS. 

It is straightforward to analytically obtain the energy eigenvalues of the MBS Hamiltonian, which read:
\bea
E&=&0,\pm \sqrt{R_1}\qquad{\rm with}\qquad\label{e1}\no\\
R_1&=&\left(t_{11}^{ab}\right)^2C_{ab}^2+\left(t_{11}^{bc}\right)^2S_{bc}^2+\left(t_{11}^{ca}\right)^2C_{ca}^2\,.\label{e1e2}
\eea

\noi The eigenvalue $E=0$ indicates that one MBS is always present. Furthermore, $R_1$ is always nonzero except at the high-symmetry point $(\pi,\pi)$ where both $C_{ab}$ and $C_{ca}$ are vanishing. As mentioned above, there, the three MBSs belong to the same chirality sector. In fact, the emergence of the point node at P can be also understood by means of a $\pi$-Berry phase (i.e. a $\mathbb{Z}_2$ topological invariant) as we discuss in Sec.~\ref{sec:Topology}. Hence, we find that the number of MBSs in the (111)-junction is either one or three.

\begin{figure*}[t!]
\begin{center}
\includegraphics[width=17.0cm]{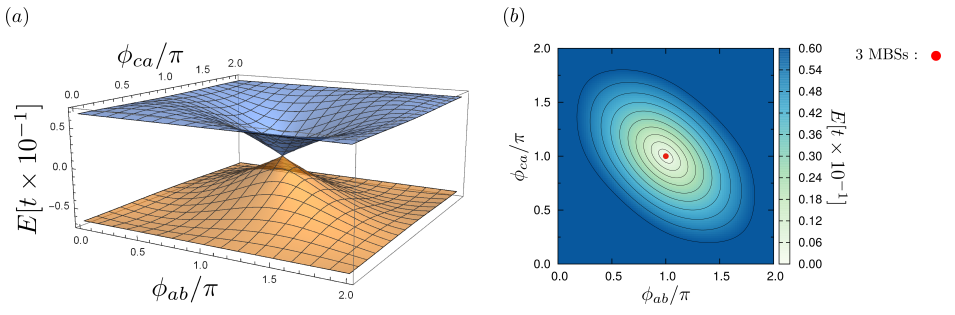}
\caption{(a) ABS energy spectra for the (111)-junction as a function of the superconducting phase differences $\phi_{ab}$ and $\phi_{ca}$. In (b) we show the energy-contour map of the states with positive energy, i.e. $E>0$. At the junction's interface, one zero-energy state always persists independently of the applied phase differences, thus giving rise to a Majorana flat band in the two-dimensional synthetic phase-difference space $(\phi_{ab},\phi_{ca})$. For graphical clarity we omit this flat band in the ABS spectra reported in the panel (a). Notably, a point node with three MBSs occurs at $(\phi_{ab},\phi_{ca})=(\pi,\pi)$.}
\label{fig:111}
\end{center}
\end{figure*}

\subsection{(112)-junction}\label{sec:112}

The outcomes of our real-space BdG analysis conducted for the three-terminal (112)-junction depicted in Fig.~\ref{fig:config}(b), are presented in Figs.~\ref{fig:112}(a) and~(b). Four MBSs appear at the intersection for zero dot-TSC tunnel couplings. When the tunnel couplings are switched on, we do not find any ZES flat bands, which implies that all the energy eigenvalues depend on $\phi_{ab}$ and $\phi_{ca}$ and, thus, the MBSs hybridize into two standard fermionic degrees of freedom giving rise to four ABS levels. Two of these ABS dispersions yield a Dirac cone-like structure centered at $(\pi,\pi)$, while the two remaining ABS energy solutions exhi\-bit a more complex type of dispersion in the phase-difference space. These properties become evident from Fig.~\ref{fig:112}(b). We identify the appearance of line band-touchings with two MBSs at $\phi_{ca}=\pi$ and $\phi_{ab}+\phi_{ca}=2\pi$. 

These features are understood by considering the following low-energy Hamiltonian based on the coupled MBS appearing near the intersection, i.e.:
\begin{align}
\widehat{\cal H}_{112}=\left(\begin{array}{cccc}
0&-it^{ab}_{11}C_{ab}&-it^{ca}_{11}C_{ca}
&it^{ca}_{12}C_{ca}\\
it^{ab}_{11}C_{ab}&0&it^{bc}_{11}S_{bc}&-it^{bc}_{12}S_{bc}\\
it^{ca}_{11}C_{ca}&-it^{bc}_{11}S_{bc}&0&0\\
-it^{ca}_{12}C_{ca}&it^{bc}_{12}S_{bc}&0&0
\end{array}\right)\,.
\label{h112}
\end{align}

\begin{figure*}[t!]
\begin{center}
\includegraphics[width=17.0cm]{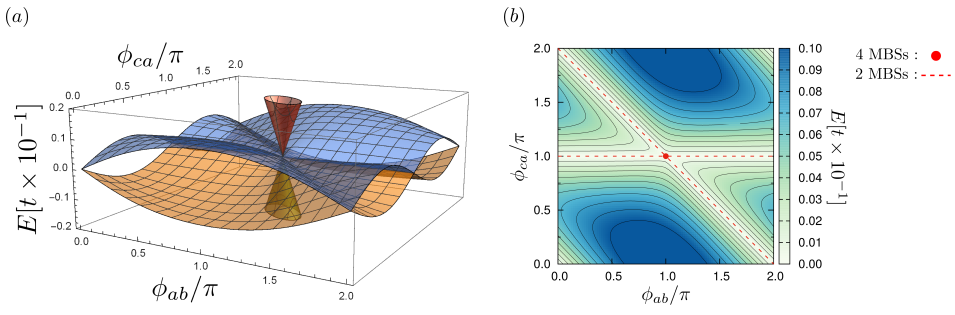}
\caption{(a) ABS energy spectra for the (112)-junction in terms of the superconducting phase differences $\phi_{ab}$ and $\phi_{ca}$. In (b) we report the energy-contour map of the states with positive energy, i.e. $E>0$. Four MBSs (red point) occur at the high-symmetry point $(\phi_{ab},\phi_{ca})=(\pi,\pi)$. Nodal lines (red dashed) with two MBSs are obtained along the symmetry directions $\phi_{ca}=\pi$ and $\phi_{ab}+\phi_{ca}=2\pi$.}
\label{fig:112}
\end{center}
\end{figure*}

\noi We point out that the two MBSs appearing on the edge of the $c$-th TSC do not couple directly because they belong to the same chirality sector. This becomes explicitly manifest by the vanishing matrix elements at positions (3,4) and (4,3) in Eq.~\eqref{h112}. This of course holds under the assumption that the $c$-th TSC posseses chiral symmetry also after contacting the quantum dot. The eigenvalues are analytically obtained, and read:
\bea
E&=&\pm E_{\pm}\quad{\rm with}\quad
E_{\pm}=\sqrt{\dfrac{R_2\pm\sqrt{R^2_2-4V_2}}{2}}\,,
\label{pme1}\\
R_{2}&=&\left(t_{11}^{ab}\right)^2C_{ab}^2+\left[\left(t_{11}^{bc}\right)^2+\left(t_{12}^{bc}\right)^2\right]S_{bc}^2\no\\
&&+\left[\left(t_{11}^{ca} \right)^2+\left(t_{12}^{ca}\right)^2\right]C_{ca}^2\,, \\
V_{2}&=&\big(t_{12}^{bc}t_{11}^{ca}-t_{11}^{bc}t_{12}^{ca}\big)^2 C_{ca}^2 S_{bc}^2\,. 
\label{beta2}
\eea

\noi The energy bands $\pm E_-$ touch at the symmetry point $\mathrm{P}$, as well as along the symmetry lines
\bea
\phi_{ca}=\pi\phd{\rm and}\phd\phi_{ab}+\phi_{ca}=2\pi\,,\label{l2}
\eea

\noi since, there, we have $V_2=0$. Notably, the latter condition also reflects the vanishing of the Pfaffian of $\widehat{\cal B}_{112}=i\widehat{\cal H}_{112}$, which reads: ${\rm Pf}(\widehat{\cal B}_{112})=C_{ca}S_{bc}\big(t_{11}^{bc}t_{12}^{ca}-t_{12}^{bc}t_{11}^{ca}\big)$.

\begin{figure*}[t!]
\begin{center}
\includegraphics[width=17.0cm]{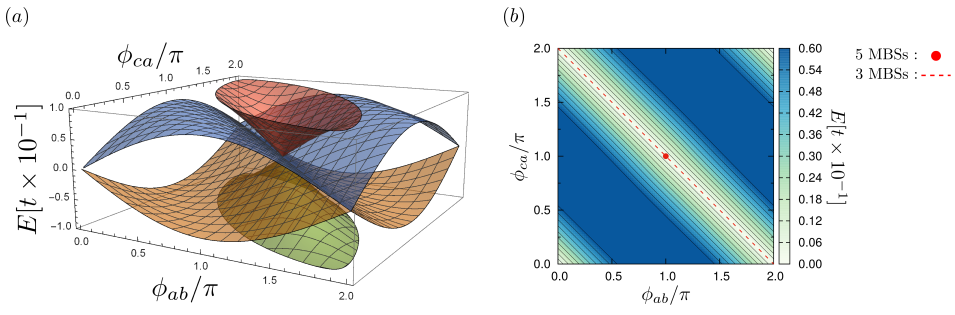}
\caption{(a) ABS energy spectra for the (122)-junction as a function of the phase differences $\phi_{ab}$ and $\phi_{ca}$ appearing between the respective superconducting leads. In (b) we report the energy-contour map of the states with positive energy, i.e. $E>0$. Five MBSs (red point) occur at the high-symmetry P point with synthetic-space coordinates $(\phi_{ab},\phi_{ca})=(\pi,\pi)$. Nodal lines (red dashed) with two MBSs are obtained along the symmetry directions. At the junction's interface, one zero-energy state always persists independently of the applied phase differences, thus giving rise to a flat band in the two-dimensional synthetic phase-difference space $(\phi_{ab},\phi_{ca})$. For graphical clarity we omit this flat band in the ABS spectra reported in the panel (a).}
\label{fig:122}
\end{center}
\end{figure*}

Along the nodal line $\phi_{ca}=\pi$, the two edge MBSs of the $c$-th TSC and the edge MBS of the $a$-th TSC belong to the same chirality sector of $\hat{\Gamma}(\phi_c)$. Therefore, according to the earlier-mentioned property (ii), the edge MBS of the $b$-th TSC can couple to only a single linear combination made up from the three MBSs arising from the $a$-th and $c$-th TSCs. As a collateral consequence, two MBSs remain uncoupled and, thus, lead to two zero eigenvalues. The presence of two MBSs can be alternatively explained by analyzing the rank of the low-energy Hamiltonian in Eq.~\eqref{h112}. Specifically, the rank decreases from $4$ to $2$ when $\phi_{ca}=\pi$ and, consequently, there are two ZESs.

The remaining nodal-line condition $\phi_{ab}+\phi_{ca}=2\pi$ is equivalent to $\phi_c=2\pi+\phi_b$. Now, it is the three MBSs ori\-gi\-na\-ting from the $b$-th and $c$-th TSCs that belong to the same chirality sector. When this takes place, the (2,3), (2,4), (3,2) and (4,2) elements of the matrix in Eq.~\eqref{h112} are zero. The condition $\phi_{ab}+\phi_{ca}=2\pi$ also leads to the reduction of rank in Eq.~(\ref{h112}) from $4$ to $2$. Thus, two MBSs persist, since the MBS of the $a$-th TSC couples to only one linear combination constructed from the three remaining MBSs. The presence of the line nodes are explained in terms of a $\mathbb{Z}_2$ topological number and extra symmetries in the parameter space whose details are presented in Sec.~\ref{sec:Topology}. Eqs.~\eqref{pme1} and~\eqref{beta2} also indicate that two ABS remain at zero energy when the condition $t_{12}^{bc}t_{11}^{ca}=t_{11}^{bc}t_{12}^{ca}$ holds. We have numerically confirmed that two MBSs stay uncoupled irrespectively of the va\-lues that $\phi_{ab}$ and $\phi_{ca}$ take. However, such zeros are only accidental. Nonetheless, together with the latter case, we conclude that in the most general case, the number o MBSs in the (112)-junction can be either $0$, $2$ or $4$.

\begin{figure*}[t!]
\begin{center}
\includegraphics[width=\textwidth]{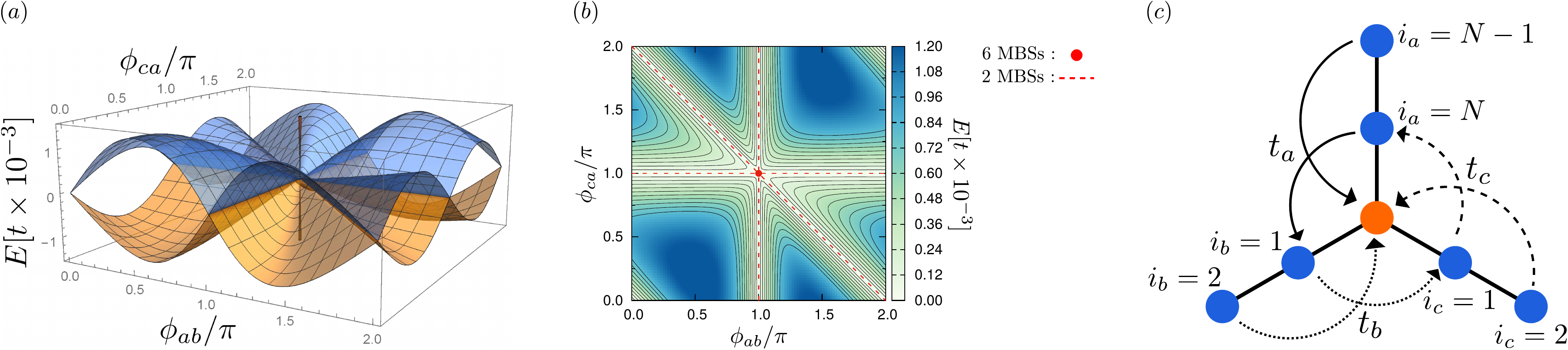}
\caption{(a) ABS energy spectra for the (222)-junction as a function of the phase differences $\phi_{ab}$ and $\phi_{ca}$. In (b) we report the energy-contour map of the states with positive energy, i.e. $E>0$. Six MBSs (red point) occur at the high-symmetry position $(\phi_{ab},\phi_{ca})=(\pi,\pi)$. Nodal lines (red dashed) with two MBSs are obtained along the symmetry directions. The spectrum with three nodal lines is obtained by including extra chiral symmetry preserving terms at the junction's interface: the next-nearest hoppings nearby the dot as shown in (c) and a Zeeman field on the dot site. Otherwise, zero-energy states typically persist independently of the applied phase differences, thus giving rise to two Majorana flat band. The spectra in (a),(b) are obtained for the following representative parameters: ${t}_a=0.7\,t$, ${t}_b=0.5\,t$, and ${t}_c=0.4\,t$, and the Zeeman field $h^{D}_x=0.1\,t$ and $h^{D}_z=0.2\,t$.  
}
\label{fig:222}
\end{center}
\end{figure*}


\subsection{(122)-junction}

Next, we discuss the (122)-junction which is depicted in Fig.~\ref{fig:config}(c). For such a configuration we obtain a maximum of five MBSs at the junction's interface. The resul\-ting ABS spectra, obtained by means of the diagonalization of the BdG Hamiltonian on the lattice, are shown in Figs.~\ref{fig:122}(a) and~(b). One MBS is always present independently of the two imposed phase differences. As we also did previously, we also here do not plot in Fig.~\ref{fig:122}(a) the corresponding flat-band dispersion for reasons of gra\-phi\-cal cla\-ri\-ty. 

Besides the Majorana flat band, we also find ABS ZESs occurring along the nodal lines defined by $\phi_{ab}+\phi_{ca}=2\pi$ (which is equivalent to $\phi_c=2\pi+\phi_b$). The two MBSs originating from the $b$-th TSC, as well as the two MBSs stemming from the $c$-th TSC possess the same chirality eigenvalue when $\phi_c=2\pi+\phi_b$. Thus, the edge MBS of the $a$-th TSC couples to a single linear combination formed by the MBSs of the $b$-th and $c$-th TSCs. Conclusively, three MBSs remain uncoupled and yield the respective ZESs along the nodal line $\phi_{ab}+\phi_{ca}=2\pi$.

Once again, the features of the ABS spectra can be well-reproduced by inferring the spectrum of the effective low-energy Hamiltonian coupling the MBSs at the intersection. This Hamiltonian reads:
\begin{align}
\widehat{\cal H}_{122}=\qquad\qquad\qquad\qquad\qquad\qquad\qquad\qquad\qquad\qquad\no\\
\left(\begin{array}{ccccc}
0&-it^{ab}_{11}C_{ab}&it^{ab}_{12}C_{ab}&-it^{ca}_{11}C_{ca}&it^{ca}_{12}C_{ca}\\
it^{ab}_{11}C_{ab}&0&0&it^{bc}_{11}S_{bc}&-it^{bc}_{12}S_{bc}\\
-it^{ab}_{12}C_{ab}&0&0&-it^{bc}_{21}S_{bc}&it^{bc}_{22}S_{bc}\\
it^{ca}_{11}C_{ca}&-it^{bc}_{11}S_{bc}&it^{bc}_{21}S_{bc}&0&0\\
-it^{ca}_{12}C_{ca}&it^{bc}_{12}S_{bc}&-it^{bc}_{22}S_{bc}&0&0
\end{array}\right)\,.
\label{h122}
\end{align}

\noi The two MBSs residing on either the $b$-th or the $c$-th TSC do not couple to each other since they are protected by chiral symmetry. This property is reflected in the pre\-sen\-ce of two $2\times2$ zero-element matrix blocks in the corresponding $b$- and $c$-MBS subspaces of the Hamiltonian matrix. The eigenvalues of the latter take the form:
\bea
E=0,\pm E_{\pm}\quad{\rm with}\quad E_{\pm}=\sqrt{\dfrac{R_3\pm\sqrt{R^2_3-4V_3}}{2}}\,,\qquad\label{pme2}
\eea

\noi where, similar to previous sections, we have introduced the parameters
\bea
R_3&=&\left[\left(t_{11}^{ab}\right)^2+\left(t_{12}^{ab} \right)^2\right]C_{ab}^2+\left[\left(t_{11}^{ca}\right)^2+\left(t_{12}^{ca}\right)^2\right]C_{ca}^2\no\\
&&+\left[\left(t_{11}^{bc}\right)^2+\left(t_{12}^{bc}\right)^2+\left(t_{21}^{bc}\right)^2+\left(t_{22}^{bc}\right)^2\right]S_{bc}^2\,,
\eea

\bea 
V_3&=&S_{bc}^2\left[\big(t_{21}^{bc}t_{11}^{ab}-t_{11}^{bc}t_{12}^{ab}\big)^2C_{ab}^2+\big(t_{22}^{bc}t_{11}^{ab}-t_{12}^{bc}t_{12}^{ab}\big)^2C_{ab}^2\right.\no\\ 
&&+\big(t_{11}^{bc}t_{22}^{bc}-t_{12}^{bc}t_{21}^{bc}\big)^2S_{bc}^2+\big(t_{12}^{bc}t_{11}^{ca}-t_{11}^{bc}t_{12}^{ca}\big)^2C_{ca}^2\no\\
&&\left.+\big(t_{22}^{bc}t_{11}^{ca}-t_{21}^{bc}t_{12}^{ca}\big)^2 C_ {ca}^2\right]\,.
\eea

In the present case, the nodal line appears when $V_3$ and, in turn, $E_-$ vanish. This takes place for $S_{bc}=0$, and results in three ZESs. On the other hand, the energy bands $\pm E_+$ touch each other at the point ${\rm P}$, which is in agreement with the numerical results depicted in Fig.~\ref{fig:122}(a). We thus conclude that the number of MBSs in a (122)-junction is either one, three or five.\\

\subsection{(222)-junction}
\textcolor{black}{
Finally, we discuss the behavior of MBSs in (222)-junction shown in Fig.~\ref{fig:config}(d). Numerical results obtained by the analysis on the lattice indicate that the behavior of the ABSs depends on the character of the chiral symmetry conserving terms at the junction's interface. Indeed, for the electronic processes indicated in Eq.~\eqref{h_link}, two MBSs survive irrespectively of the values chosen for the phase differences among the three TSCs. On the other hand, as described in Fig.~\ref{fig:222}(a) and~(b), the inclusion of next-nearest neighbor hoppings and a non-vanishing spin-polarization on the dot generally lead to an ABS spectrum with dispersive modes and three nodal lines. These features can be well explained by considering the effective low-energy Hamiltonian for the (222) configuration:
\begin{widetext}
\bea
\widehat{\cal H}_{222}=\left(\begin{array}{cccccc}
0&0&-it^{ab}_{11}C_{ab}&it^{ab}_{12}C_{ab}&-it^{ca}_{11}C_{ca}&it^{ca}_{12}C_{ca}\\
0&0&it^{ab}_{21}C_{ab}&-it^{ab}_{22}C_{ab}&it^{ca}_{21}C_{ca}&-it^{ca}_{22}C_{ca}\\
it^{ab}_{11}C_{ab}&-it^{ab}_{21}C_{ab}&0&0&it^{bc}_{11}S_{bc}&-it^{bc}_{12}S_{bc}\\
-it^{ab}_{12}C_{ab}&it^{ab}_{22}C_{ab}&0&0&-it^{bc}_{21}S_{bc}&it^{bc}_{22}S_{bc}\\
it^{ca}_{11}C_{ca}&-it^{ca}_{21}C_{ca}&-it^{bc}_{11}S_{bc}&it^{bc}_{21}S_{bc}&0&0\\
-it^{ca}_{12}C_{ca}&it^{ca}_{22}C_{ca}&it^{bc}_{12}S_{bc}&-it^{bc}_{22}S_{bc}&0&0
\end{array}\right)\,.
\label{h222}
\eea
{\color{black}Let us first consider the case when two MBSs stay decoupled. Indeed, when the following relations hold:
\begin{align}
t^{\alpha \beta}_{11}=t^{\alpha \beta}_{22}, \quad t^{\alpha \beta}_{12}=t^{\alpha \beta}_{21},
\label{eq:222hop}
\end{align}
the eigenvalues of $\widehat{\cal H}_{222}$ can be calculated analytically, and read:}  
\bea
E&=&0,0,\pm \sqrt{E_{\pm}}\,,\label{Rubc}\\
E_{\pm}&=&\left(t_{11}^{ab}\pm t_{12}^{ab}\right)^2C_{ab}^2+\left(t_{11}^{bc}\pm t_{12}^{bc}\right)^2S_{bc}^2+\left( t_{11}^{ca} \pm t_{12}^{ca} \right)^2 C_{ca}^2\,.
\eea
The two zero eigenvalues suggest the existence of flat bands independently of the precise values of the other parameters. 
%
%
\indent Although it is not easy to obtain the generic expression for the ABS spectra analytically, the degeneracy of the ZESs can be checked in a different way. One can for instance investigate the behavior of ${\rm Pf}(\widehat{\cal B}_{222})$, which reads:
\bea
{\rm Pf}(\widehat{\cal B}_{222})&=&C_{ab}C_{ca}S_{bc} T_{222}\,, \\
T_{222}&=& 
t^{ab}_{11}\big(t^{bc}_{22}t^{ca}_{21}-t^{ca}_{22}t^{bc}_{21}\big)+t^{ca}_{12}\big(t^{ab}_{21}t^{bc}_{21}-t^{ab}_{22}t^{bc}_{11}\big)
+t^{ab}_{12}\big(t^{ca}_{22}t^{bc}_{11}-t^{ca}_{21}t^{bc}_{12}\big)+t^{ca}_{11}\big(t^{ab}_{22}t^{bc}_{12}-t^{ab}_{21}t^{bc}_{22}\big)\,.
\label{eq:Pf222}
\eea
\end{widetext}
It is immediate to check that $T_{222}=0$ as long as Eq.~(\ref{eq:222hop}) is satisfied.
When $T_{222}\neq 0$ can be realized, however, the structure of the above expression suggests the appea\-ran\-ce of only point and line nodes in the ABS spectra. This is what we found by modifying the Hamiltonian with the inclusion of the next-neighbor hopping near the intersection's point and a Zeeman field ($h^{D}$) in the $xz$ plane on the dot as shown in Fig.~\ref{fig:222}(c). The ABS spectra as a function of the applied phase differences are displayed in Figs.~\ref{fig:222}(a) and~(b). There exist three line nodes along high symmetry directions that cross at the point ${\rm P}=(\pi,\pi)$. We argue that these extra terms at the interface can mimic a low energy configuration such as $T_{222}\neq 0$. 
In such a case, the number of MBSs can be zero, two, or six in the space of the phase differences between the superconducting leads.\\
\indent It is interesting to point out that for the 222 configuration the engineering of the interface coupling between the TSCs can turn flat MBSs into dispersive ones in the presence of a non-vanishing phase difference.  
}


\section{Topological Invariants}\label{sec:Topology}

In the following two paragraphs we expose a number of aspects regarding the origin of the two types of nodal ABS spectra encountered here, and the symmetries that make their presence possible. Even more, we construct suitable invariants that reflect their topological protection.

\subsection{Protection of point nodes}

The emergence of point nodes in the (111)- and (112)-junctions can be understood in a unified manner by re\-lying on the structure of a Hamiltonian describing the coupling of four MBSs. For fully-gapped or nodal-points-containing ABS spectra stemming from four coupled MBSs $\gamma_{1,2,3,4}$, arranged in a multicomponent Majorana spinor as $\bm{\Gamma}^{\intercal}=(\gamma_1\,\,\gamma_2\,\,\gamma_3\,\,\gamma_4)$, one can decompose $\widehat{\cal B}$ into two $\mathfrak{so}(3)$ algebras, i.e. $i\widehat{\cal B}=\sum_{n=1,2}\bm{g}_n\cdot\bm{J}_n$. See also Refs.~\onlinecite{Houzet18,KotetesPRL2019}. By employing the Pauli matrices $\bm{\lambda}$ and $\bm{\kappa}$, the $\mathfrak{so}(3)$ generators read $\bm{J}_1=\big(\lambda_1\kappa_2,\,\lambda_2,\,\lambda_3\kappa_2\big)/2$ and $\bm{J}_2=\big(\lambda_2\kappa_1,\,\kappa_2,\,\lambda_2\kappa_3\big)/2$. Therefore, we find: 
\bea
\bm{g}_1&=&-\big(B_{14}-B_{23},\,B_{13}+B_{24},\,B_{12}-B_{34}\big)\,,\\
\bm{g}_2&=&-\big(B_{14}+B_{23},\,B_{12}+B_{34},\,B_{13}-B_{24}\big)\,.
\eea

In the case of the (112)-junction, it is straightforward to obtain the respective two $\bm{g}$ vectors:
\bea
\bm{g}_1&=&\big(t_{12}^{ca}C_{ca}-t_{11}^{bc}S_{bc},\,-t_{12}^{bc}S_{bc}-t_{11}^{ca}C_{ca},\,-t_{11}^{ab}C_{ab}\big),\quad\phd\\
\bm{g}_2&=&\big(t_{12}^{ca}C_{ca}+t_{11}^{bc}S_{bc},\,-t_{11}^{ab}C_{ab},\,t_{12}^{bc}S_{bc}-t_{11}^{ca}C_{ca}\big).\quad
\eea

\noi It is convenient to perform the shift $(\phi_{ab},\phi_{ca})\mapsto(\phi_{ab},\phi_{ca})-(\pi,\pi)$, which by virtue of the $4\pi$-periodicity of the MBS couplings, it effects the transformation $(C_{ab},C_{ca},S_{bc})\mapsto(S_{ab},S_{ca},-S_{bc})$ and thus places the high-symmetry point P at the origin of the coordinate system. Therefore, the respective vectors read:
\bea
\bm{g}_1'&=&\big(t_{12}^{ca}S_{ca}+t_{11}^{bc}S_{bc},\,t_{12}^{bc}S_{bc}-t_{11}^{ca}S_{ca},\,-t_{11}^{ab}S_{ab}\big),\quad\phd\label{eq:Gs112a}\\
\bm{g}_2'&=&\big(t_{12}^{ca}S_{ca}-t_{11}^{bc}S_{bc},\,-t_{11}^{ab}S_{ab},\,-t_{12}^{bc}S_{bc}-t_{11}^{ca}S_{ca}\big).\quad\label{eq:Gs112b}
\eea

\noi In the shifted coordinate system, these vector satisfy $\bm{g}_{1,2}'(-\phi_{ab},-\phi_{ca})=-\bm{g}_{1,2}'(\phi_{ab},\phi_{ca})$. The structure of the above $\bm{g}'$ vectors is similar to the one obtained for the generally-warped helical surface of topological insulators which contain a single Dirac point and are characterized by a $\pi$-Berry phase~\cite{piBerryPhase1,piBerryPhase2}. Therefore, a similar $\mathbb{Z}_2$ invariant characterizes the point node in the present case. 

While in topological insulators the vanishing of the $\bm{g}$ vectors at the high-symmetry point is imposed by time-reversal symmetry, in the present situation the two Dirac points are a consequence of chiral symmetry. Thus, we expect the violation of chiral symmetry to lead to the opening of a spectral gap at P, since in this event, P no longer constitutes a high-symmetry point of the synthetic phase-difference space.

The above approach can be extended to the (111)-junction. For this purpose, one formally adds an auxi\-lia\-ry MBS in order to take advantage of the above results which relied on four coupled MBSs. For instance, this auxiliary MBS may be one of the MBSs which are located far away from the junction, and are uncoupled from the interfacial MBSs. According to the above procedure, the relevant $\bm{g}$ vectors for the (111)-junction are obtained from Eqs.~\ref{eq:Gs112a} and~\ref{eq:Gs112b} by setting $t_{12}^{ca}=t_{12}^{bc}=0$, and read:
\bea
\bm{g}_1'&=&\big(+t_{11}^{bc}S_{bc},\,-t_{11}^{ca}S_{ca},\,-t_{11}^{ab}S_{ab}\big),\\
\bm{g}_2'&=&\big(-t_{11}^{bc}S_{bc},\,-t_{11}^{ab}S_{ab},\,-t_{11}^{ca}S_{ca}\big).
\eea

\noi From these two equations we infer that $|\bm{g}_{1}'|=|\bm{g}_2'|$, which leads to the twofold-degenerate ZES correspon\-ding to the nonlocal fermionic degree of freedom consi\-sting of one MBS away from the junction and the single uncoupled MBS always present at the junction.

We conclude this paragraph by demonstrating how to apply the above to the case of the (222)-junction, where two Dirac points appear. Since the two bands leading to the line nodes are irrelevant for the present discussion, we consider their flat-band limit. This is achieved by setting the parameter values $t^{\alpha\beta}_{11}=t^{\alpha\beta}_{22}$ and $t^{\alpha\beta}_{12}=t^{\alpha \beta}_{21}$ ($\forall\alpha,\beta=a,b,c$). In this case, the Hamiltonian commutes with the matrix $\mathds{1}_3\otimes\kappa_1$. This property allows us to block-diagonalize it into two irreducible sectors corresponding to the eigenstates of $\mathds{1}_3\otimes\kappa_1$, that we here label using the quantum number $\kappa=\pm1$. Thus, the two Hamiltonian blocks read:
\begin{align}
\widehat{\cal H}_{222}^\kappa=\left(\begin{array}{ccc}
0&-i\big(t^{ab}_{11}+\kappa t^{ab}_{12}\big)C_{ab}&-i\big(t^{ca}_{11}+\kappa t^{ca}_{12}\big)C_{ca}\\
&0&i\big(t^{bc}_{11}+\kappa t^{bc}_{12}\big)S_{bc}\\
&{\rm h.c.}&0
\end{array}\right)\,.
\end{align}

\noi In analogy to the procedure followed for the (111)-junction, one here augments the number of MBS per block by one, and defines the respective $\bm{g}_{\kappa}$ vectors. As mentioned above, the introduction of these vectors allows for the $\mathbb{Z}_2$ topological classification of the protected point nodes at P. 

\subsection{Protection of line nodes}

Here, we focus on the origin of the line nodes in the $(\phi_{ab},\phi_{ca})$ synthetic space and prove that they can also be characterized by a topological index. To see this, we start from the low-energy effective model Hamiltonian given in Eq.~(\ref{h1}), which is expressed in terms of the MBSs which are localized near the intersection of the Y-junction. As mentioned earlier, the anticommutation relations satisfied by the MBS operators impose that the matrix part of the above Hamiltonian is skew-symmetric, i.e. $\widehat{\cal H}^\intercal=-\widehat{\cal H}$, and one can introduce the real skew-symmetric matrix $\widehat{\cal B}=i\widehat{\cal H}$, for which, the relation $\det[\widehat{\cal B}(\phi_{ab},\phi_{ca})]={\rm Pf}[\widehat{\cal B}(\phi_{ab},\phi_{ca})]^2$ holds. Therefore, within this low-energy model, the nodes can be determined by searching the solutions of $\det[\widehat{\cal B}(\phi_{ab},\phi_{ca})]=0$ or equi\-va\-lently the ones of ${\rm Pf}[\widehat{\cal B}(\phi_{ab},\phi_{ca})]=0$. We distinguish two cases depending on whether the rank of $\widehat{\cal B}$ is even or odd, since in the latter case the Pfaffian of $\widehat{\cal B}$ is trivially zero. 

When the rank of the effective Hamiltonian is even, the Pfaffian becomes zero only for particular values of the phase differences. Specifically, when a line node emer\-ges in the $(\phi_{ab},\phi_{ca})$ plane, the sign of ${\rm Pf}[\widehat{\cal B}(\phi_{ab},\phi_{ca})]$ changes across the line node. Therefore, we introduce the following $\mathbb{Z}_2$ topological index~\cite{Kobayashi,Zhao,Bzdusek} for the line node:
\bea
{\cal N}={\rm sgn}\left\{{\rm Pf}\big[\widehat{\cal B}(\phi_{ab},\phi_{ca})\big]{\rm Pf}\big[\widehat{\cal B}(\phi_{ab}',\phi_{ca}')\big]\right\}\,,\label{eq:topo}
\eea

\noi where ${\cal N}=\pm1$ denotes the absence (existence) of a line node in a path connecting $(\phi_{ab},\phi_{ca})$ to $(\phi_{ab}',\phi_{ca}')$. 

In contrast, when the rank of the matrix Hamiltonian $\widehat{\cal H}$ is odd, the Pfaffian of the related matrix $\widehat{\cal B}$ matrix is zero and, thus, we cannot directly employ it for inferring the topological properties of the spectrum. Nevertheless, the skew-symmetric cha\-racter of $\widehat{\cal H}(\phi_{ab},\phi_{ca})$ implies that it is dictated by an odd number of ZESs. Therefore, by effecting a unitary transformation on $\widehat{\cal H}(\phi_{ab},\phi_{ca})$, we can project out the trivial ZES subspace and define a skew-symmetric matrix of an even rank, for which, a $\mathbb{Z}_2$ topological invariant as in Eq.~\eqref{eq:topo} is well defined. 

The geometrical structure, location and open character of the nodal lines are imposed by the presence of chiral symmetry and the invarianc of $\widehat{\cal B}$ under $2\pi$ phase shifts. The latter are reflected in Eq.~\eqref{eq:relUalpha}. Given the choice $\phi_{bc}=-(\phi_{ab}+\phi_{ca})$, the inva\-rian\-ce of the system under $2\pi$-shifts in $\phi_{a,b,c}$ leads to the following symmetry relations when $2\pi$-shifts for the phase differences $\phi_{ab,ca}$ are considered: 
\bea
\hat{U}_1^\dag\widehat{\cal B}(\phi_{ab},\phi_{ca})\hat{U}_1&=&\widehat{\cal B}(\phi_{ab}+2\pi,\phi_{ca}),\label{eq:relU1}\\
\hat{U}_2^\dag\widehat{\cal B}(\phi_{ab},\phi_{ca})\hat{U}_2&=&\widehat{\cal B}(\phi_{ab},\phi_{ca}+2\pi),\label{eq:relU2}\\
\hat{U}_{12}^\dag\widehat{\cal B}(\phi_{ab},\phi_{ca})\hat{U}_{12}&=&\widehat{\cal B}(\phi_{ab}+2\pi,\phi_{ca}+2\pi),\quad\label{eq:relU1U2}
\eea 

\noi where after Eq.~\eqref{eq:relUalpha}, we have $\hat{U}_1=\hat{U}_a\hat{U}_c$, $\hat{U}_2=\hat{U}_c$ and $\hat{U}_{12}=\hat{U}_a$.

Using the above, we now explicitly discuss how Eq.~\eqref{eq:topo}, in conjuction with chiral symmetry and the relations of Eqs.~\eqref{eq:relU1}-\eqref{eq:relU1U2}, predicts the presence of line nodes in the synthetic $(\phi_{ab},\phi_{ca})$ space for the case of a (112)-junction. For such a junction, $\hat{U}_1=\mathrm{diag}(-1,1,-1,-1)$, $\hat{U}_2=\mathrm{diag}(1,1,-1,-1)$, and $\hat{U}_{12}=\mathrm{diag}(-1,1,1,1)$. By means of the relation ${\rm Pf}(\widehat{U}^\intercal\widehat{\cal B}\hat{U})=\det(\widehat{U}){\rm Pf}(\widehat{\cal B})$, we have:
\bea
{\rm Pf}\big[\widehat{\cal B}_{112}(\phi_{ab}+2\pi,\phi_{ca})\big]&=&-{\rm Pf}\big[\widehat{\cal B}_{112}(\phi_{ab},\phi_{ca})\big],\quad\phd\ph\label{eq:PfU1}\\
{\rm Pf}\big[\widehat{\cal B}_{112}(\phi_{ab},\phi_{ca}+2 \pi)\big]&=&+{\rm Pf}\big[\widehat{\cal B}_{112}(\phi_{ab},\phi_{ca})\big],\quad\phd\ph\label{eq:PfU2}\\
{\rm Pf}\big[\widehat{\cal B}_{112}(\phi_{ab}+2\pi,\phi_{ca}+2\pi)\big]&=&-{\rm Pf}\big[\widehat{\cal B}_{112}(\phi_{ab},\phi_{ca})\big]\quad\phd\ph\label{eq:PfU1U2}
\eea

\noi where we used $\det(\hat{U}_2)=-\det(\hat{U}_1)=1$. Equations~\eqref{eq:PfU1} and~\eqref{eq:PfU1U2} already imply that line nodes are accessible in the ABS spectrum, since ${\rm Pf}(\widehat{\cal B}_{112})$ changes sign. Moreover, the $2\pi$ ($4\pi$) periodicity in the phase $\phi_{ca}$ ($\phi_{ab}$), implies that $\phi_{ca}=\pi$ ($\phi_{ab}=\pi$) is (not) ne\-ces\-sa\-ri\-ly a high-symmetry line. Nonetheless, the $4\pi$ pe\-rio\-di\-ci\-ty in $\phi_{ab}$ results in the equivalence $\phi_{ab}=-\pi\equiv3\pi$. Moreover, chiral symmetry imposes that P$(\pi,\pi)$ is a high-symmetry point, i.e. an inversion center sati\-sfying $(\pi,\pi)\equiv (-\pi,-\pi)$. This further renders the line $\phi_{ab}+\phi_{ca}=2\pi$ a high-symmetry line. In fact, the nodal lines coincide with the ho\-ri\-zon\-tal axis $\phi_{ca}=\pi$, and the diagonal $\phi_{ab}+\phi_{ca}=2\pi$. Specifically, the horizontal nodal line is directly obtai\-na\-ble by exploiting that P is an inversion-symmetric point and $\phi_{ca}=\pi\equiv-\pi$, since Eq.~\eqref{eq:PfU1} yields: $-{\rm Pf}\big[\widehat{\cal B}_{112}(\phi_{ab},\pi)\big]={\rm Pf}\big[\widehat{\cal B}_{112}(\phi_{ab}+\pi+\pi,\pi)\big]\equiv{\rm Pf}\big[\widehat{\cal B}_{112}(\phi_{ab},-\pi)\big]\equiv{\rm Pf}\big[\widehat{\cal B}_{112}(\phi_{ab},\pi)\big]$. Si\-mi\-lar arguments establish the diagonal nodal line of Fig.~\ref{fig:112}, since once again the presence of the inversion center P implies that ${\rm Pf}[\widehat{\cal B}_{112}(\phi_{ab},2\pi-\phi_{ab})]=0$, through the combination of Eqs.~\eqref{eq:PfU1} and~\eqref{eq:PfU2}.

As we showed above, in the (112)-junction the nodal lines result from the invariance of the Hamiltonian under $2\pi$ shifts and the concomitant $4\pi$-periodic dependence of ${\rm Pf}[\widehat{\cal B}_{112}(\phi_{ab},\phi_{ca})]$ on one of the phase differences. However, this is not the case for the (222)-junction, where ${\rm Pf}[\widehat{\cal B}_{222}(\phi_{ab},\phi_{ca})]$ is $2\pi$ periodic on both arguments due to the even number of MBSs that each TSC supports before contact. This is reflected in the following form that Eqs.~\eqref{eq:relU1}-\eqref{eq:relU1U2} take for the (222)-junction:
\bea
{\rm Pf}\big[\widehat{\cal B}_{222}(\phi_{ab}+2\pi,\phi_{ca})\big]&=&{\rm Pf}\big[\widehat{\cal B}_{222}(\phi_{ab},\phi_{ca})\big],\quad\phd\ph\label{eq:Pf222U1}\\
{\rm Pf}\big[\widehat{\cal B}_{222}(\phi_{ab},\phi_{ca}+2 \pi)\big]&=&{\rm Pf}\big[\widehat{\cal B}_{222}(\phi_{ab},\phi_{ca})\big],\quad\phd\ph\label{eq:Pf222U2}\\
{\rm Pf}\big[\widehat{\cal B}_{222}(\phi_{ab}+2\pi,\phi_{ca}+2\pi)\big]&=&{\rm Pf}\big[\widehat{\cal B}_{222}(\phi_{ab},\phi_{ca})\big].\quad\phd\ph\label{eq:Pf222U1U2}
\eea

\noi For this junction configuration, chiral symmetry establishes $\phi_{ca}=\pi$, $\phi_{ab}=\pi$ and $\phi_{ab}+\phi_{ca}=2\pi$ as high-symmetry lines of the synthetic space. In fact, as it is immediately discernable from Eq.~\eqref{eq:Pf222}, nodal lines appear at these three high-symmetry lines. The emergence of the nodal lines can be understood by additional symmetry properties that ${\rm Pf}[\widehat{\cal B}_{222}(\phi_{ab},\phi_{ca})]$ possesses along the high-symmetry lines. Specifically, along $\phi_{ab}=\pi$ and $\phi_{ab}=2\pi-\phi_{ca}$ ($\phi_{ca}=\pi$) one finds $\widehat{\cal B}_{222}(\phi_{ca}+2\pi)=-\widehat{\cal B}_{222}(\phi_{ca})$ ($\widehat{\cal B}_{222}(\phi_{ab}+2\pi)=-\widehat{\cal B}_{222}(\phi_{ab})$). Since for a skew-symmetric $2N\times2N$ matrix the respective Pfaffian is an $N$-th order polynomial of the matrix entries, we respectively find ${\rm Pf}[\widehat{\cal B}_{222}(\phi_{ca}+2\pi)]=-{\rm Pf}[\widehat{\cal B}_{222}(\phi_{ca})]$ (${\rm Pf}[\widehat{\cal B}_{222}(\phi_{ab}+2\pi)]=-{\rm Pf}[\widehat{\cal B}_{222}(\phi_{ab})]$). These relations render the $\mathbb{Z}_2$ topological indices defined in Eq.~\eqref{eq:topo} nontrivial which, in turn, stabilize the nodal lines. 

\begin{figure*}[t!]
\begin{center}
\includegraphics[width=0.24\textwidth]{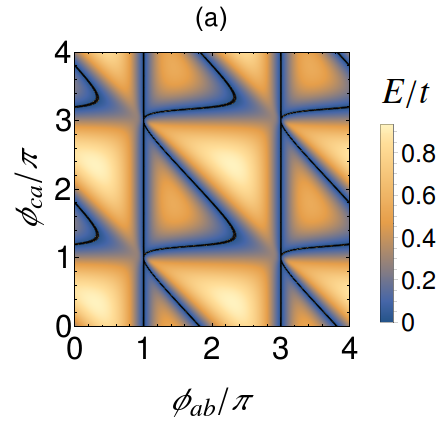}
\includegraphics[width=0.24\textwidth]{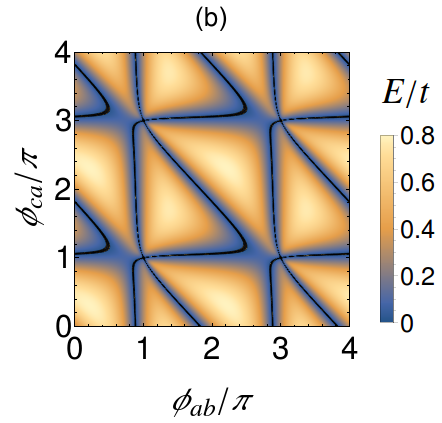}
\includegraphics[width=0.24\textwidth]{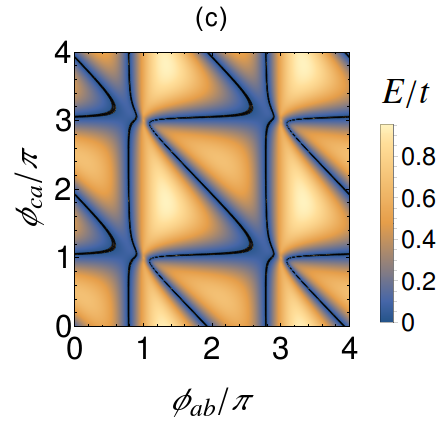}
\includegraphics[width=0.24\textwidth]{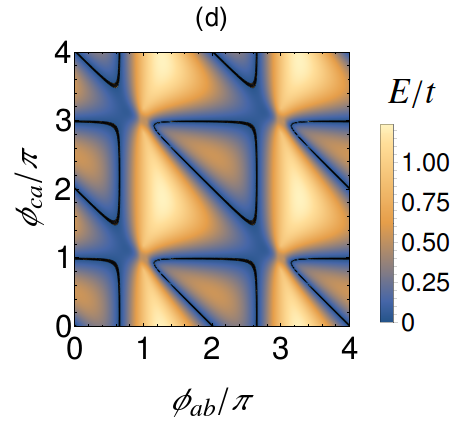}
\caption{Density plots of the lowest energy Andreev bound state in the $(\phi_{ab},\phi_{ac}$) phase-difference plane for the (222)-junction in the presence of hybridization terms $t_j$ (with $j=a,b,c$ labelling each superconducting lead). We show few representative cases for $t_j$ smaller than the tunneling energy scale $t$ to highlight the evolution of the nodal lines. This energy scale hierarchy is expected when the chiral symmetry protecting the multiple MBSs on a given TSC is weakly violated. The black lines mark the place of the nodal lines where the ensuing Pfaffian vanishes, i.e. ${\rm Pf}(\widehat{\cal B}_{222})=0$. In (a) we report a representative physical configuration with one nonvanishing $t_j$ amplitude (i.e. $t_c=0.5 t$, $t_a=t_b=0$) demonstrating the persistence of a single high-symmetry nodal line and the simultaneous emergence of a nodal-chain structure. The presence of two nonzero hybridization amplitudes (i.e. $t_a=0.6t$, $t_b=0$, $t_c=0.5t$) does not remove the nodal chain which, however, now develops away from high-symmetry lines, as shown in (b). For two nonvanishing hybridization terms a suitable choice of the parameters can also give an isolated point node at the high-symmetry position ($\pi,\pi)$ together with closed loops. When all the MBS hybridization channels are present, one can have open lines  (see (c) for $t_a=0.6t$, $t_b=t_c=0.5t$) and closed  loops (see (d) for $t_a=t_b=0.9t$, $t_c=0.5t$). The other parameters in the Hamiltonian $\widehat{\cal H}_{222}$ have the following amplitudes: $t_{11}^{ab}=t_{22}^{ab}=5t$, $t_{12}^{ab}=t$, 
$t_{11}^{ca}=t_{22}^{ca}=t_{11}^{bc}=t_{22}^{bc}=2t$, $t_{21}^{ab}=t_{12}^{ca}=t_{21}^{ca}=t_{12}^{bc}=t_{21}^{bc}=4t$. }
\label{fig:222CSB1}
\end{center}
\end{figure*}

\section{Effects of MBS hybridization on a given topological superconductor}

In the previous sections, we considered that chiral symmetry is preserved before and after the three TSC leads become coupled through the linking quantum dot, thus, generally allowing for protected multiple MBSs per edge. Here, we relax this assumption and investigate the effects of the violation of chiral symmetry in every TSC, already before they get contacted. Since for the present discussion we are not particularly interested in the pre\-ci\-se source of the chiral-symmetry violation, we examine its consequences from a qualitative point of view, and consider nonzero couplings between the MBSs appearing on the same edge of a given TSC. We remark at this point that we consider that the Josephson-junction link does not introduce any further chiral-symmetry brea\-king terms. When such a scenario takes place, the spectrum can be rendered fully-gapped for all values of the two independent phase differences. Note, however, that the Majorana flat bands are inert to the chiral-symmetry breaking terms, since they originate from the presence of an odd number of MBSs near the interface, with one of them always remaining uncoupled.

Introducing these chiral-symmetry violating couplings between two MBS of a given TSC, immediately yields an important conclusion regarding the stability of the nodal spectra. That is, in the presence of chiral-symmetry breaking terms, nodal points are generally unstable while nodal lines are robust and protected by the PHS and FP conservation of the MBS Hamiltonian. Specifically, the point node at P in the (111)- and (112)-junction cases is removed for an arbitrarily weak strength of a chiral-symmetry breaking perturbation. In contrast, the two Dirac point nodes appearing in the (122)- and (222)-junctions can be both removed only when chiral-symmetry vio\-la\-ting couplings are introduced to all pairs of MBSs arising from the same TSC. 

As mentioned above, nodal lines are instead robust, and can be removed only after a certain threshold for the strength of the chiral-symmetry violating terms has been reached. Until that point, the addition of chiral-symmetry vio\-la\-ting perturbations mainly affects the geo\-me\-tri\-cal structure of the nodal lines. Specifically, these no longer need to be open or appear along high-symmetry lines. There exist two possible scenarios that allow for the removal of the nodal lines. The first takes place when pairs of open curved nodal lines of identical shape can meet and annihilate. However, this requires certain fine tuning. The second and most prominent mechanism occurs by means of a Lifshitz transition~\cite{Lifshitz}. In this case, the nodal topology changes and the line evolves from open to a closed loop, which can now continuously shrink to a point upon varying the parameters and get lifted from zero energy. Notably, the topological index of Eq.~\eqref{eq:topo} remains employable even when chiral symmetry is broken and the nodal lines are not open. However, predicting the location and shape of the nodal lines is no longer straightforward.

A numerical evaluation of the ABS energy spectrum when chiral symmetry is violated, unveils a rich landscape of nodal lines, including nodal loops and nodal chains, which were recently experimentally discovered in the band structure of a number of nodal semimetallic materials~\cite{Liu,Liu1,Neupane,Bian,Wu,Wang,Yan}. The energy spectra for a number of representative chiral-symmetry vio\-la\-ting configurations are depicted in Figs.~\ref{fig:222CSB1} and~\ref{fig:222CSB2}. In the former (latter) the chiral-symmetry breaking terms have strengths which are quite smaller (larger) than the tunneling energy scales. The consideration of these two limits is useful for reproducing two different physical situations. The weak chiral-symmetry breaking limit corresponds to the case where pairs of MBSs on a given TSC hybridize into tri\-vial nonzero energy ABSs. Nevertheless, as long as the symmetry brea\-king terms are weak, one expects the underlying topological nature of the MBSs to be still reflected in the structure of the ABS spectrum, e.g. in a similar fashion to the $4\pi$-periodic Josephson effect~\cite{Kitaev01} obtained from two coupled MBSs. In the antipodal limit, a chiral-symmetry breaking term that strongly hybridizes two MBSs on a given TSC, can be equivalently viewed as the hybridization between two MBSs which comprise a trivial ABS that lacks of a topological origin and/or protection~\cite{Kells,CXLiu,Vuik,Moore}. In this case, the hybridization term \textit{is not} associated with the violation of chiral symmetry. 

Experimentally, the presence of such a trivial ABS may be mistaken for two topologically-protected MBSs ori\-gi\-na\-ting from a single chiral-symmetry preserving TSC. This is because, these trivial ABSs may even lie at zero energy, thus reproducing the same spectra that two chiral-symmetry protected MBSs would do. However, such a situation can only happen in a very restricted region of the parameter space. More importantly, as mentioned above, the hybridization of two MBSs for\-ming a trivial ABS is unrelated to chiral-symmetry vio\-la\-tion and, thus, it is nonzero also when this symmetry is preserved. Remarkably, one can rely on this property to develop experimental strategies that can allow disentangling the two scenarios. 

\begin{figure*}[t!]
\begin{center}
\includegraphics[width=0.24\textwidth]{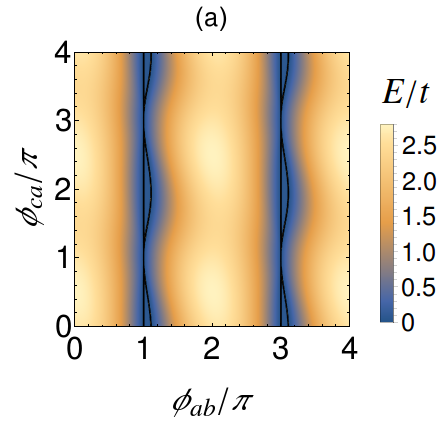}
\includegraphics[width=0.24\textwidth]{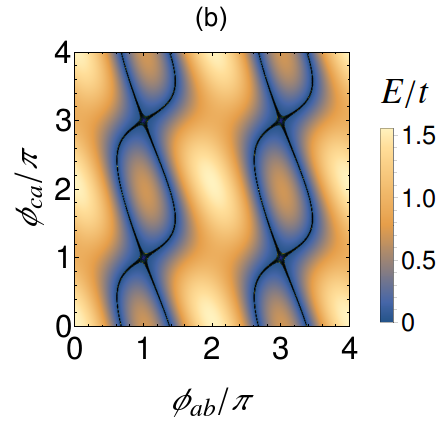}
\includegraphics[width=0.24\textwidth]{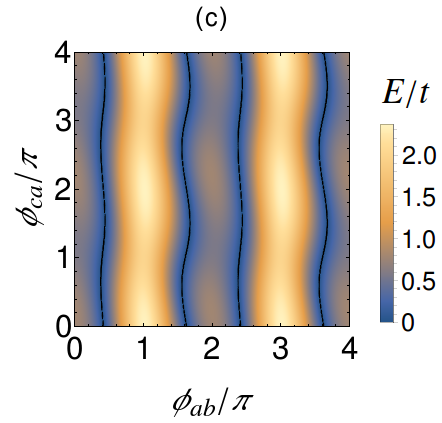}
\includegraphics[width=0.24\textwidth]{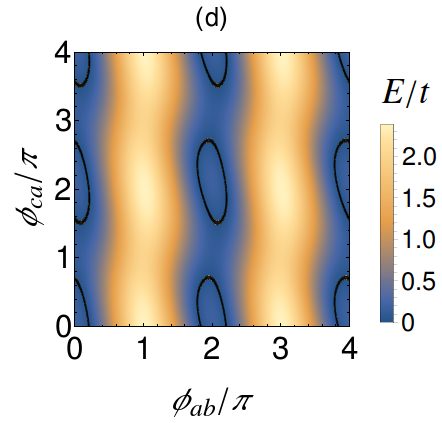}
\caption{Density plots of the lowest energy Andreev bound state in the $(\phi_{ab},\phi_{ac}$) phase-difference plane for the (222)-junction in the presence of hybridization terms $t_j$ (with $j=a,b,c$ labelling each superconducting lead). The black lines mark the place of the nodal lines where the ensuing Pfaffian vanishes, i.e. ${\rm Pf}(\widehat{\cal B}_{222})=0$. Here, we report on the case of strong MBS hybridization corresponding to $t_j$ being much larger than the tunneling energy scale $t$. Such a situation may describe the presence of a trivial ABS on the edge of a given superconducting lead instead of two coupled topologically-protected MBSs. In (a) we show the case of only one nonzero hybridization term (e.g. $t_a=t_b=0$ and $t_c=10t$). We find a nodal-chain structure which is stable even for very large values of $t_j$. Notably, it is concentrated near $\phi_{ab}=\pi$, thus possessing a synthetic-space profile which is indicative of the presence of a trivial ABS. In (b) we have robust nodal chains concentrated near $\phi_{ab}=\pi$, even when assuming that one hybridization term is zero (e.g. $t_a=10t$, $t_b=0$, $t_c=15t$). In contrast, when all the hybridization terms are nonvanishing with the $t_j$ amplitudes exceeding a critical value, the nodal chains first evolve into the open lines shown in panel (c) and, then, they turn into the loops depicted in panel (d). Finally, they disappear when the loops shrink into points upon increasing the strength of the hybridization terms. In (c) the parameters are $t_a=6t$, $t_b=2t$, $t_c=25t$ while in (d) $t_a=9t$, $t_b=2t$, $t_c=25t$.}
\label{fig:222CSB2}
\end{center}
\end{figure*}

This appears possible by experimentally controlling a chiral-symmetry preserving knob, e.g. the strength and/or orientation of the magnetic field appearing in Eq.~\eqref{h_op}. In particular, if the observed nodal ABS spectra originate from topologically-protected MBSs, which may still additionally experience weak local chiral-symmetry violating perturbations, the spectrum and other related signatures should remain unchanged as long as the mo\-di\-fi\-ca\-tion of the strength of the knob in question does not effect a topological phase transition. In stark contrast, modifying the strength of a chiral-symmetry preserving field will immediately affect the Andreev spectrum when a trivial ABS is present. Even more, if this knob can be manipulated so that the hybridization term between the MBSs comprising the trivial ABS dominates over the tunnel coupling energy scales, one finds an ABS spectrum which is clearly distinct to the one obtained from topologically-protected MBSs.

Representative ABS energy spectra in the strong MBS hybridization limit are depicted in Fig.~\ref{fig:222CSB2}. For a (112)-junction, one finds that the nodal line is now concentrated about the $\phi_{ab}=\pi$ axis, instead of coinci\-ding with the high-symmetry lines $\phi_{ab}+\phi_{ca}=2\pi$ and $\phi_{ca}=\pi$ which are obtained when chiral-symmetry and topologically-protected MBSs are present. For a (122)-junction, one is required to experimentally dif\-fe\-ren\-tia\-te among the topologically-nontrivial case and situations with one or two trivial ABSs appearing at the $b$-th and/or $c$-th TSC. If a trivial ABS is located at the $c$-th ($b$-th) TSC, the nodal line spectra is centered about the line $\phi_{ab}=\pi$ ($\phi_{ca}=\pi$). On the other hand, when two ABSs are present and the hybridization energy scales are sufficiently strong, the resulting spectra are fully-gapped. This is after excluding the single Majorana flat band which is always present for such an interface. Therefore, similar to the (112)-junction, the presence of tri\-vial ABSs yields nodal spectra which are concentrated at lines which are otherwise not accesssible in the respective topological configuration. 

A similar clear-cut behavior takes place for the (222)-junction where one, two or three trivial ABSs become possible. For three trivial ABSs one finds fully-gapped spectra which should be easily detactable in experiments. For one (two) ABSs one obtains a nodal line (point at P) when chiral symmetry is preserved. In more detail, when the trivial ABS is associated with the two MBSs of the \{$a$, $b$, $c$\}-th TSC, a nodal line spectrum will be found centered about the line \{$\phi_{ab}+\phi_{ca}=2\pi$, $\phi_{ca}=\pi$, $\phi_{ab}=\pi$\}, respectively. 

\textcolor{black}{\section{Josephson transport and nodal line spectra}
The analysis of the previous sections has demonstrated that nodal line spectra are generally robust and, therefore, they are the most prominent features to be observed in experiments. In the present section, we provide few key elements of the Josephson-transport in three-terminal devices with nodal line spectra. The recommended experimental strategy for probing the nodal line spectra is to measure the Josephson current which is ge\-ne\-ra\-ted when sweeping the phase along lines in the 2D synthetic space. For this purpose it is required to adiabatically vary the two superconducting phase differences $\phi_{ab}$ and $\phi_{ca}$, a task which can be experimentally achieved by threading magnetic fluxes through suitable loops of the device.\\
{\color{black}\indent A crucial factor that determines the resulting Josephson current profile, is whether the fermion parity (FP) of the three-terminal junction is conserved during such a transport experiment. When FP is conserved the even and odd FP sectors correspond to different energies, and thus give rise to different Josephson responses. As a consequence, the Josephson current is continuous and, depending on the configuration, it may be $4\pi$- or $2\pi$-periodic~\cite{Kitaev01,FuKaneJ,CarloJ}. In contrast, when FP is not preserved, the resulting Josephson current originates from all the occupied ABSs and is $2\pi$ periodic. In the following, we focus on the most general case of broken FP, and demonstrate that the presence of nodal lines is still experimentally identifiable through discontinuities marking the Josepshon current.\\}
\indent To compute the Josephson effect for the Hamiltonian on the lattice, one can employ the formulation in Refs.~\cite{furusaki,asano:prb2001}, where the Josephson current {\color{black}flowing through the $\alpha$-th} TSC is expressed by
\begin{align}
J_{\alpha}=\frac{iet}{\hbar}T\sum_{\omega_n}\mathrm{Tr}\left[\hat{\mathcal{G}}_\alpha(j_0,j_0+1;\omega_n)-\hat{\mathcal{G}}_\alpha(j_0+1,j_0;\omega_n)\right],\label{Green}
\end{align}
where $\hat{\mathcal{G}}_\alpha$ is the Matsubara Green's function {\color{black}describing the $\alpha$-th} TSC, and $\omega_n$ denote the respective Matsu\-ba\-ra frequencies. The current $J_{\alpha}$ flows from the $\alpha$-th TSC to the link dot. For the generic three-terminal junction, the calculation of the Josephson current is performed by inserting three normal lattice sites, with each acting as a bridge between the given TSC and the dot. Here, $j_0$ indicates those normal lattice sites.\\ 
\indent Before presenting the results, it is quite instructive to start with the (111) case, which can guide the analysis for the other cases. For this configuration the Josephson current can be calculated analytically within the scat\-te\-ring formalism. The results at zero temperature are given by
\bea
J_a&=&\frac{e\Delta}{\hbar}\frac{|t_n|(\sin\phi_{ab}+\sin\phi_{ac})}
{\sqrt{\cos^2\left(\frac{\phi_{ab}}{2}\right)+\cos^2\left(\frac{\phi_{ca}}{2}\right)+\sin^2\left(\frac{\phi_{bc}}{2}\right)}}\,,\quad\\
J_b&=&\frac{e\Delta}{\hbar}\frac{|t_n|(\sin\phi_{ba}-\sin\phi_{bc})}
{\sqrt{\cos^2\left(\frac{\phi_{ab}}{2}\right)+\cos^2\left(\frac{\phi_{ca}}{2}\right)+\sin^2\left(\frac{\phi_{bc}}{2}\right)}}\,,\quad\\
J_c&=&\frac{e\Delta}{\hbar}\frac{|t_n|(\sin\phi_{ca}-\sin\phi_{cb})}
{\sqrt{\cos^2\left(\frac{ \phi_{ab} }{2}\right)+\cos^2\left(\frac{\phi_{ca}}{2}\right)+\sin^2\left(\frac{\phi_{bc}}{2}\right)}}\,,\quad
\eea
where $|t_n|^2$ is the transmission probability from one TSC to another in the normal state, which is set equal to $4/9$ for the remainder. The current conservation law implies $J_a+J_b+J_c=0$. It is possible to verify that the same expressions by exploiting the effective low-energy model Hamiltonian and {\color{black}defining the current flowing through each TSC as the result of the derivative of the system's energy with respect to the phase imposed on the given TSC.} Indeed, the supercurrent flowing through each TSC can be obtained by means of the equation
\bea
J_\alpha=\frac{e}{\hbar}\frac{\partial}{\partial\phi_\alpha}(-\sqrt{R_1}), 
\eea
where $-\sqrt{R_1}$ corresponds to the energy dispersion of the occupied states below zero energy as given in Eq.~\eqref{e1e2} with $t_{11}^{\alpha,\alpha^\prime}=|t_n|$ for all $\alpha$ and $\alpha^\prime$. The relation $\phi_{ab}+\phi_{bc}+\phi_{ca}=0$ should be taken into account after differentiating the energy with respect to the phases. In general, one can demonstrate that the relations:
\bea
J_\alpha=\frac{e}{\hbar}\sum_{\lambda} \frac{\partial}{\partial \phi_\alpha} E_{\lambda}, 
\label{eq:j-phi}
\eea
hold true for all three-terminal junctions at zero tem\-pe\-ra\-tu\-re, where $E_\lambda<0$ define the energies of the occupied states indicated by $\lambda$. Eq.~\eqref{eq:j-phi} provides a general relationship between the current and the electronic state energy in multi-terminal Josephson junctions. We verify that the outcome of the analysis in Eq.~\eqref{Green} is consistent with that obtained by the analysis of the energy spectra. 
\\
\indent At this point, to highlight the relationship between the Josephson current and the nodal structure in the 2D synthetic space, we present in Fig~\ref{fig:112_j} a number of representative results for the (112)-junction. We start by con\-si\-de\-ring the Josephson current as a function of $\phi_{ca}$ for fixed $\phi_{ab}=\pi/2$. The results in (a), (b) and (c) correspond with the currents flowing in each arm of the junction, i.e. $J_a$, $J_b$ and $J_c$, respectively. As shown in Fig.~\ref{fig:112} at $\phi_{ab}=\pi/2$, the nodal lines occur at $\phi_{ca}=\pi$ and $3\pi/2$. We find that $J_a$ exhibit a discontinuity only at $\phi_{ca}=\pi$. The same holds for $J_b$ with a jump only at $\phi_{ca}=3\pi/2$. Instead, $J_c$ has a jump both at $\phi_{ca}=\pi$ and $\phi_{ca}=3\pi/2$. Such discontinuity stems from the energy dispersion near the nodal lines. In the vicinity of $\phi_{ca}=\pi$, the ABS ener\-gy dispersion of Eq.~(\ref{pme1}) is expanded as:
\begin{figure}[t!]
\begin{center}
\includegraphics[width=0.99\columnwidth]{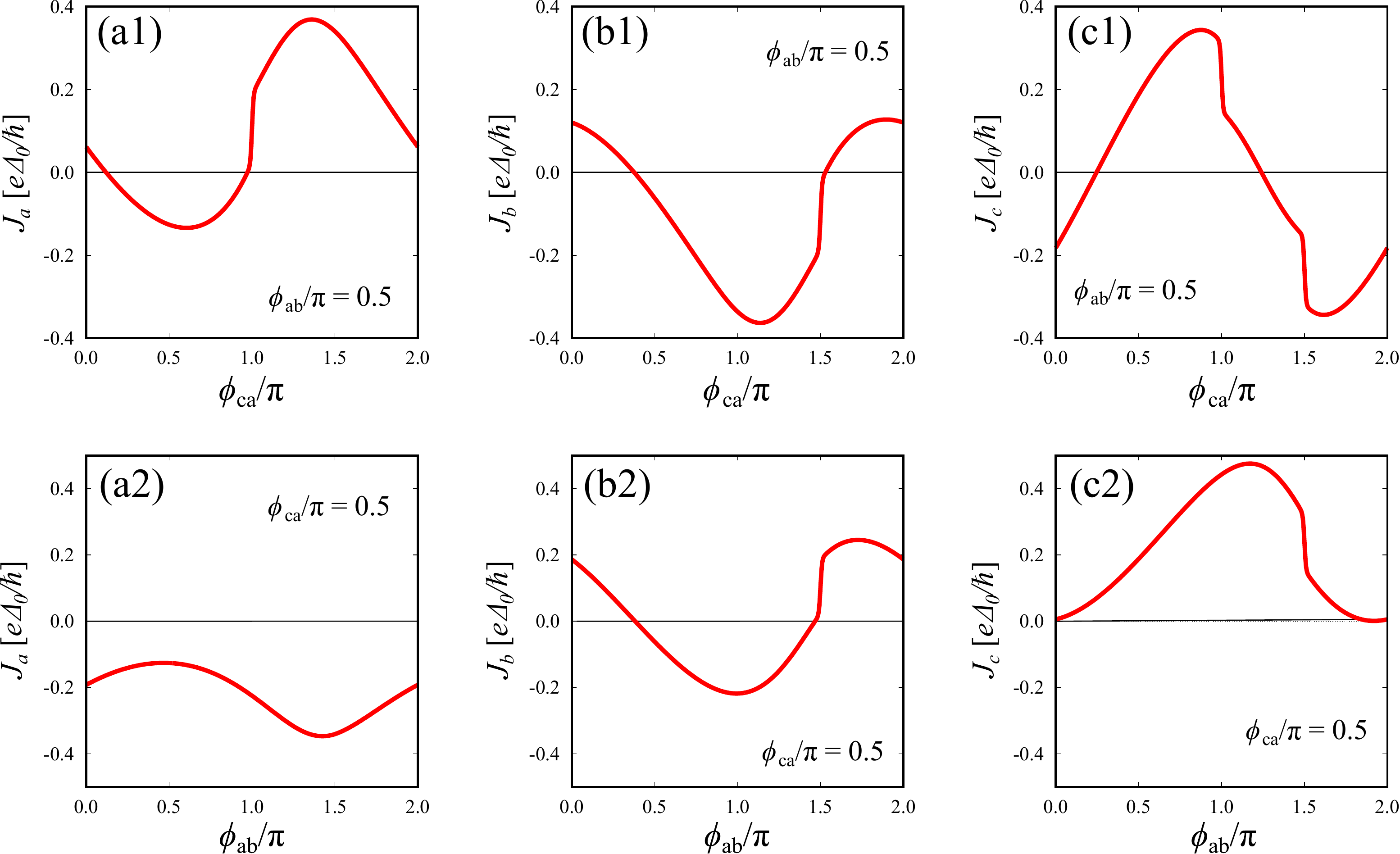}
\caption{The Josephson current in the three TSCs. 
The currents at $\phi_{ab}=\pi/2$ in $a$-, $b$- and $c$-th TSC are plotted as a function of $\phi_{ca}$ in (a1), (b1) and (c1), respectively. The currents at $\phi_{ca}=\pi/2$ in $a$-, $b$- and $c$-th TSC are plotted as a function of $\phi_{ab}$ in (a2), (b2) and (c2), respectively.  
}
\label{fig:112_j}
\end{center}
\end{figure}
\bea
-E_-\approx-\epsilon_0\,|\phi_{ca} - \pi|\,,
\eea
with $\epsilon_0>0$. The current can be then calculated as
\bea
J_\alpha \approx \frac{e}{\hbar}\left[\frac{\partial}{\partial\phi_\alpha}(-E_+)-\epsilon_0 \,\mathrm{sgn}(\phi_{ca}-\pi)\right]\,.
\eea
The second term is discontinuous at the nodal line. As shown in Fig.~\ref{fig:112_j}, the contribution to the current from the $-E_+$ band renders the current-phase relationship discontinuous. 
\\
\indent Similar trends are obtained for a phase-difference sweep in $\phi_{ab}$ while setting $\phi_{ca}=\pi/2$. 
As expected, the current $J_b$ (Fig.~\ref{fig:112_j}(b2)) and $J_c$ (Fig.~\ref{fig:112_j}(c2)) show the jump at $\phi_{ca}=3\pi/2$ because of the presence of the nodal line depicted in Fig.~\ref{fig:112}(b). The results clearly suggest that the current-phase relationship at low temperature well reflects the nodal structures of the
ABS energy dispersions. {\color{black}Finally, we note that our conclusions can be directly generalized to the remaining junction confi\-gu\-ra\-tions and for other paths in the synthetic phase space.}
}

\section{Conclusions}

We study a three-terminal Josephson junction which consists of one-dimensional topological superconductors (TSCs) with multiple chiral-symmetry protected MBSs. For our exploration we consider all the possible cases with 1 or 2 MBSs per TSC. Our main focus is on the band topology of the ABS spectra in the synthetic phase-difference space, and the evolution of the number of MBSs occurring at the junction's intersection. Table~\ref{Table:ABSspectraSummary} summarizes the general nodal properties of the ABS spectra with respect to the MBSs, i.e. it provides information about the type of nodal line and/or nodal point confi\-gu\-ra\-tion arising, and the number of uncoupled Majorana flat bands present.

The analysis relies on both (i) a real-space numerical approach incorporating the microscopic details of the p-wave superconducting nanowires forming the Y-junction, and (ii) an effective low-energy model describing the hybridization of the zero-energy MBSs located at the intersection of the Y-junction. In this manner, our results have a generic character and are applicable to TSCs other than p-wave superconductors, Specifically, including engineered TSCs based on superconductor-semiconductor hybrids~\cite{Jay,Jens,Haim,KlinovajaTRIparafermions,KotetesNoZeeman,Hell,PientkaPlanar,HaimReview,KotetesPRL2019,Heimes2,JinAn,Silas,Andolina} and topological magnetic chains~\cite{Heimes2,JinAn,Silas,Andolina}, where the multiple MBSs at each TSC edge arise from the presence of either a Kramers degeneracy or a sublattice symmetry.

One of the main outcomes of our work is that the Andreev spectrum defined in the synthetic space of the two independent phase dif\-fe\-ren\-ces developing among the three TSCs, has a structure that depends nontrivially on the number of MBSs characterizing each one of the TSCs comprising the three-terminal junction. The same conclusion also applies to the resulting Josephson currents. We find that when the junction consists of TSCs with an unequal number of MBSs, then there always e\-xist dif\-fe\-rent types of nodal lines in the Andreev spectra. These lines result from the invariance of the low-energy Hamiltonian upon 2$\pi$-shifts in the phase differences, a symmetry property which imposes sign changes of the related Pfaffian in the synthetic space. These nodal lines are robust against the application of a phase difference across two of the three nanowires or in a locking phase mode where the relative phases across each pair of superconducting leads is varied at the same time. Remarkably, the structure of the line nodes in the synthetic space differs depending on whether one has the $(112)$, $(122)$ or $(222)$ topological configuration in the Y-junction. This aspect determines the periodicity of the current with respect to the phase differences when the total fermion parity is preserved, or when the latter is not, it controls the possible emergence of discontinuities in the current when driving the phase differences along directions in the 2D synthetic space. 

In more detail, we find that the additional presence of a single MBS when moving from the $(111)$ to the $(112)$ configuration leads to a significant changeover in the energy-phase relation of the Andreev spectra. Dirac points give their place to nodal lines with direct effects and consequences on the transport properties of the multi-terminal heterostructure. Similarly, by switching from the $(112)$ to $(122)$ topological configuration, one can modify the structure of the nodal lines in the synthetic space, thus affecting the phase-drive dependence of the Josephson transport properties. Another interesting feature of the examined topological Y-junction is provided by the fact that in the $(112)$ and $(122)$ configurations, there are two MBSs occurring along the nodal lines that coexist with other two MBSs at the point node at the high-symmetry point P$(\pi,\pi)$ in the phase difference space. This implies that, by a suitable phase control, one may realize a sort of Majorana box with a tunable number of zero energy modes, thus potentially opening perspectives towards the design of topologically-protected qubit or qudit configurations. 

We also discuss the robustness and stability of the nodal spectra against perturbations which hybridize pairs of MBSs defined for a given TSC. These can be either attributed  to the violation of the chiral symmetry which protects the two MBSs of the TSC, or to the presence of trivial ABSs. In the latter case, the hybridization is unrelated to the violation of chiral symmetry and is present even when this is preserved. Motivated by this aspect, we put forward strategies to experimentally differentiate between the two scenarios. Specifically, we show that tuning a chiral-symmetry preserving knob can uncover the possible presence of trivial ABSs. Our analysis additionally reveals that nodal line spectra are more stable than nodal points against external perturbations that hybridize local pairs of MBSs. 

We conclude with discussing the expected behavior of the Josephson response from nodal line spectra. 
The geometrical shape of the nodal lines inferred from the Josephson current response for given phase trajectories in the synthetic space, can be used to assess the topological configuration of the Y-junction, as well as the degree of the chiral-symmetry violation or possible presence of trivial ABSs. Finally, based on the presented results, one can also reverse-engineer distinct Josephson transport properties by controllably violating the chiral-symmetry on one or more TSC leads. 
  
\section*{Note Added}

While the present manuscript was under review in Physical Review B, the preprint of Ref.~\onlinecite{Houzet19} appeared. This work has negligible overlap with ours and discusses the possibility of a quantized transconductance in three-terminal junctions consisting of TSCs harboring a single MBS per edge, which corresponds to a (111)-junction in our notation.  

\begin{acknowledgments}
This work was supported by ``Topological Materials Science'' (Nos. JP15H05852 and JP15K21717) from the Ministry of Education, Culture, Sports, Science and Technology (MEXT)
of Japan, JSPS Core-to-Core Program (A. Advanced Research Networks), Japanese-Russian JSPS-RFBR project (Nos.~2717G8334b and 17-52-50080). S. K. was supported by the CREST project (JPMJCR16F2) from Japan Science and Techno\-lo\-gy Agency (JST), and the Building of Consortia for the Development of Human Resources in Science and Te\-chno\-lo\-gy. M.C. acknowledges support by the project "Two-dimensional Oxides Platform for SPIN-orbitronics na\-no\-techno\-logy (TOPSPIN)" funded by the MIUR Progetti di Ricerca di Rilevante Interesse Nazionale (PRIN) Bando 2017 - Grant 20177SL7HC.
\end{acknowledgments}

\appendix

\section{Majorana bound state wave functions}\label{app:A}

In this section we provide details on the wave functions of the MBSs. The four energy eigenvalues of Eq.~\eqref{h_op} are given by $\pm\sqrt{[\epsilon(k)\pm|\boldsymbol{h}|]^2+\Delta^2(k)}$. Considering the left edge of a given TSC, the bound states at energy $E$ can be described by a wave function of the form
\begin{align}
&F_L(x)
=+\left(\begin{array}{c}h_x\Delta\\ -h_- \Delta \\ h_x \omega^\ast \\ -h_- \omega^\ast \end{array}\right)e^{ik_- x} A_1
+\left(\begin{array}{c}h_x\Delta\\ -h_+ \Delta \\ h_x \omega^\ast \\ -h_+ \omega^\ast \end{array}\right)e^{ik_+ x} A_2
\nonumber\\
&+\left(\begin{array}{c}-h_x\Delta\\ -h_- \Delta \\ h_x \omega \\ -h_- \omega \end{array}\right)e^{-ik_-^\ast x} B_1
+\left(\begin{array}{c}-h_x\Delta\\ -h_+ \Delta \\ h_x \omega \\ -h_+ \omega \end{array}\right)e^{-ik_+^\ast x} B_2
, \label{wf_l}
\end{align}

\noi where $h_\pm=h_z\pm|\boldsymbol{h}|$ and $\omega=E+i\sqrt{\Delta^2-E^2}$. In the above we set $|\Delta(k)|=\Delta$, since the exact $k$-dependence of the amplitude of the pair potential is not crucial for the present discussion. However, the sign changes of the pair potential are important and have been already accounted for in Eq.~\eqref{wf_l}. The complex wave numbers $k_\pm(E)$ are calculated from the relation 
\begin{align}
\big[\epsilon(k_{\pm})\pm|\boldsymbol{h}|\big]^2+\Delta^2=&E^2,  
\end{align}

\noi where we choose both real and imaginary parts of $k_\pm=\kappa_{\pm}+i \kappa^\prime_{\pm}$ to be positive values. The coefficients $A_1$ and $A_2$ ($B_1$ and $B_2$) denote the amplitude of the wave function in the electron (hole) branch. It is easy to confirm from the boundary condition $F_L(0)=0$ that zero-energy MBSs are possible. The resulting wave functions for two edge MBSs are given by
\begin{align}
F_L(x)=&A_1 \, \psi_{-,1}\sin(\kappa_{-} x)e^{-\kappa_{-}^\prime x}\nonumber\\+&A_2 \, \psi_{-,2} \sin(\kappa_{+} x) e^{-\kappa^\prime_{+}x},
\end{align}

\noi with Eqs.~\eqref{f_minus},~\eqref{psi_m1},~and~\eqref{psi_m2} applying.

The wave function of the two bound states at the right edge of the TSC at $x=0$ can be represented in the same manner as
\begin{align}
F_R(x) =& 
A_1 \, \psi_{+,1} \sin(\kappa_{-} x)  e^{\kappa_{-}^\prime x} \nonumber\\ +&A_2 \, \psi_{+,2}  \sin(\kappa_{+} x) e^{\kappa^\prime_{+}x},
\end{align}
with Eqs.~(\ref{f_plus}), (\ref{psi_p1}) and (\ref{psi_p2}) applying.

To describe the wave function of a TSC in the phase of $|w|=1$, we first analyze the Hamiltonian for the spin-$\uparrow$ sector,
\begin{align}
H_\uparrow=\left(\begin{array}{cc}
\epsilon(k)-|\boldsymbol{h}| & \Delta(k)\\
\Delta(k)& -\epsilon(k)+|\boldsymbol{h}|
\end{array}\right),
\end{align} 

\noi where we assume that the Zeeman field is along the $z$ axis. The dispersion for a spin-$\downarrow$ electron is pushed to high energies and does not cross the Fermi level. The bound state at the left (right) edge is described by $f_{-, \uparrow}$ ($f_{+,\uparrow}$). The effects of $h_x$ are considered through the rotation of the wave function in spin space as
\begin{align}
e^{i \theta/2 \hat{\sigma}_2} f_{\pm, \uparrow}=\cos(\theta/2) f_{\pm, \uparrow} + \sin(\theta/2) f_{\pm, \downarrow}.
\end{align}

\noi The rhs corresponds to $\psi_{\pm, 1}$ in Eqs.~\eqref{psi_p1} and~\eqref{psi_m1}.\\

\section{Hybridization of Majorana bound states}\label{app:B}

The tunnel-coupling matrix elements that describe the hybridization of the MBSs are calculated from the overlap of the respective wave functions at the two edges of the pairwise coupled TSCs. This coupling is mediated by the quantom dot which links the three TSCs, as shown in Fig.~\ref{fig:model}(a). The hopping matrix elements in the low-energy effective Hamiltonian are calculated as follows:
\begin{align}
H_{\nu_\alpha, \nu_\beta}^{\alpha \beta}=&\frac{1}{2}\left[ e^{i\phi_\alpha\hat{\tau}_3/2}\, 
\psi_{\Gamma_\alpha, \nu_\alpha}^{(\alpha)} \right]^\dagger(-2\tilde{t} \hat{\tau}_3)
e^{i\phi_\beta\hat{\tau}_3/2}\, \psi_{\Gamma_{\beta}, \nu_\beta}^{(\beta)},\\
=&\frac{i}{2}\tilde{t}{\rm I}_{\Gamma_\alpha,\nu_\alpha;\Gamma_\beta,\nu_\beta}(\phi_\alpha-\phi_\beta,\theta_\alpha-\theta_\beta),\label{i_gt}
\end{align}

\noi where $\alpha,\beta =(a, b,$ and $c)$ label the TSCs. The amplitude of the hopping mediated by the dot is $\tilde{t}$, and depends on the spatial profile of the MBS  wave functions, as well as on the distance between the two edges. In addition, ${\rm I}_{\Gamma_\alpha,\nu_\alpha;\Gamma_\beta, \nu_\beta}$ depends on the chirality configuration of the MBSs, the relative phase difference, and the relative direction of the Zeeman fields in the two coupled TSCs. In Table~\ref{table:element}, we summarize the results for ${\rm I}_{\Gamma_\alpha,\nu_\alpha; \Gamma_\beta,\nu_\beta}$ where
\begin{align}
C_\phi=&\cos\left(\frac{\phi_\alpha-\phi_\beta}{2}\right),\quad 
S_\phi=\sin\left(\frac{\phi_\alpha-\phi_\beta}{2}\right),\\
C_\theta=& \cos\left(\frac{\theta_\alpha-\theta_\beta}{2}\right),\quad 
S_\theta=\sin\left(\frac{\theta_\alpha-\theta_\beta}{2}\right).
\end{align}
 
\begin{table}[h!]
\begin{center}
\begin{ruledtabular}
\begin{tabular}{lcccc}
\null & $\psi^{(\beta)}_{+, 1}$  & $\psi^{(\beta)}_{+, 2}$ & $\psi^{(\beta)}_{-, 1}$ & $\psi^{(\beta)}_{-, 2}$\\
\hline
$\psi^{(\alpha)}_{+, 1}$ &  $S_\phi C_\theta$ & $-S_\phi S_\theta$ & $-C_\phi C_\theta$ & $C_\phi S_\theta$\\
$\psi^{(\alpha)}_{+, 2}$ &  $S_\phi S_\theta$ & $S_\phi C_\theta$ & $ C_\phi S_\theta$ & $-C_\phi C_\theta$\\
$\psi^{(\alpha)}_{-, 1}$  &  $C_\phi C_\theta$ & $-C_\phi S_\theta$ & $S_\phi C_\theta$ & $-S_\phi S_\theta$\\
$\psi^{(\alpha)}_{-, 2}$  &  $-C_\phi S_\theta$ & $C_\phi C_\theta$ & $S_\phi S_\theta$ & $S_\phi C_\theta$ 
\end{tabular}
\end{ruledtabular}
\caption{
${\rm I}_{\Gamma_\alpha,\nu_\alpha;\Gamma_\beta, \nu_\beta}$ in Eq.~(\ref{i_gt}).
}
\end{center}
\label{table:element}
\end{table}

In the main text, the dependence of ${\rm I}_{\Gamma_\alpha,\nu_\alpha;\Gamma_\beta,\nu_\beta}$ on $\phi_\alpha-\phi_\beta$ is factorized and the remaining part is denoted by $t^{\alpha\beta}_{\nu_\alpha\nu_\beta}$. The signs of the hopping elements do not affect eigenvalues of the low-energy Hamiltonian.

\end{document}